\documentclass[12pt,preprint]{aastex}
\usepackage{epsfig}
\usepackage{psfig}

\newcommand{\chisq}{$\chi^2$}

\def\spose#1{\hbox to 0pt{#1\hss}}

\def\beq{\begin{equation}}
\def\enq{\end{equation}}

\def\lta{\mathrel{\spose{\lower 3pt\hbox{$\mathchar"218$}}
     \raise 2.0pt\hbox{$\mathchar"13C$}}}
\def\gta{\mathrel{\spose{\lower 3pt\hbox{$\mathchar"218$}}
     \raise 2.0pt\hbox{$\mathchar"13E$}}}

\newcommand{\PSRB}{PSR B1929$+$10}

\shorttitle{{A Multiwavelength Study of the Pulsar PSR B$1929+10$}}
\shortauthors{W.~Becker et al.}

\begin{document}

\title{{A Multi-wavelength study of the Pulsar PSR B$1929+10$\\ and its X-ray trail}}

\author{
Werner Becker\altaffilmark{1},
Michael Kramer\altaffilmark{2},
Axel Jessner\altaffilmark{3},
Ronald E. Taam\altaffilmark{4,}\altaffilmark{5},\\
Jian J. Jia\altaffilmark{6},
Kwong S. Cheng\altaffilmark{6},
Roberto Mignani\altaffilmark{7},
Alberto Pellizzoni\altaffilmark{8},\\
Andrea  de Luca\altaffilmark{8},
Agnieszka S{\l}owikowska\altaffilmark{9},
Patrizia A.~Caraveo\altaffilmark{8} }


\altaffiltext{1}{Max-Planck Institut f\"ur Extraterrestrische Physik, 85741 Garching bei M\"unchen, Germany}
\altaffiltext{2}{University of Manchester, Jodrell Bank Observatory, Macclesfield, Cheshire SK11 9DL, UK}
\altaffiltext{3}{Max-Planck Institut f\"ur Radioastronomie, Effelsberg, 53902 Bad M\"unstereifel, Germany}
\altaffiltext{4}{Northwestern University, Dep.~of Physics and Astronomy, 2145 Sheridan Road, Evanston, IL 60208}
\altaffiltext{5}{ASIAA/National Tsing Hua University - TIARA, Hsinchu, Taiwan}
\altaffiltext{6}{Department of Physics, University of Hong Kong, Hong Kong, China}
\altaffiltext{7}{Mullard Space Science Lab, University College London, Holmbury St Mary, Dorking, Surrey RH5 6NT}
\altaffiltext{8}{INAF, Istituto di Astrofisica Spaziale e Fisica Cosmica, via Bassini 15, I-20133 Milano, Italy}
\altaffiltext{9}{Nicolaus Copernicus Astronomical Center, Rabia\'nska 8, 87-100 Toru\'n, Poland}

{\quad\quad\quad\quad\quad\quad\quad Accepted for publication in \apj \, on March 16, 2006\\[-4ex]}

\begin{abstract}
  We report on the emission properties of \PSRB\, and its putative
  trail from a multi-wavelength study performed using optical, X-ray and radio 
  data.  XMM-Newton observations confirm the existence of the diffuse 
  emission with a trail morphology lying in a direction opposite to the 
  transverse motion of the pulsar. The trail spectrum is non-thermal 
  and produced by electron-synchrotron emission in the shock between 
  the pulsar wind and the surrounding medium. Radio data from the 
  Effelsberg 11cm radio continuum survey show an elongated feature 
  which roughly coincides with the X-ray trail. Three not fully 
  resolved radio sources seen in the NVSS survey data at 1.4 GHz 
  match with part of the elongated radio feature seen at 11cm.
  The emission properties observed from \PSRB\,are in excellent  
  agreement with a non-thermal and, thus, magnetospheric radiation 
  dominated emission scenario. The pulsar's X-ray spectrum is best 
  described by a single power law model with a photon index of 
  $2.72^{+0.12}_{-0.09}$.  A flux contribution from the thermal 
  emission of heated polar caps of at most $\sim 7\%$ is inferred 
  from a best fitting composite Planckian and power law spectral   
  model. A pure thermal emission spectrum consisting of two Planckian
  spectra is regarded as unlikely. A broken power law spectral model
  with $E_{break}= 0.83^{+0.05}_{-0.03}\,\mbox{keV}$ and the
  photon-indexes $\alpha_1=1.12^{+0.02}_{-0.03}$ and $\alpha_2
  =2.48^{+0.08}_{-0.07}\,$ can describe the optical and X-ray data
  entirely in terms of a non-thermal magnetospheric origin.
  The X-ray pulse profile observed in the $0.2-10$ keV band is found
  to be markedly different from the broad sinusoidal pulse profile
  seen in the low statistic ROSAT data. Fitting Gaussians to the X-ray
  light curve indicates the possible existence of three pulse
  components. A small narrow pulse, characterized by energies greater
  than 1 keV, is found to lead the radio main pulse by $\sim
  20^{\circ}$. The fraction of pulsed photons in the $0.2-10$ keV band
  is $32 \pm 4 \%$.  For the sub-bands $0.2-1.0$ keV and $1.0-2.1$ keV
  the pulsed fraction is $24 \pm 5 \%$ and $44 \pm 6\%$, respectively,
  indicating a mild energy dependence at a $\sim 2\sigma$ level.
  Simulations in the framework of an outer-gap emission model are able
  to reproduce the observed X-ray pulse profile and its phase shift
  relative to the radio pulse.

\end{abstract}


\section{INTRODUCTION \label{intro}}

 Currently, a consistent scenario for the evolution of the X-ray
 emission properties of aging rotation-powered pulsars does not 
 exist yet. This surprising fact is largely due to the lack of sufficient
 observational data. Young\footnote{In standards of high energy astronomy 
 rotation-powered pulsars are called young, middle aged and old if their 
 spin-down age is of the order of few times $10^3-10^4$ yrs, $10^5-10^6$ 
 yrs and $\ge 10^6$ yrs, respectively.  This classification is diffuse,
 though, with a smooth transition in between the different groups.} and 
 middle aged neutron stars, which  emit strong pulsed non-thermal and/or 
 surface hot-spot plus cooling emission, were studied reasonably well in 
 the X-ray band. In contrast, until recently, most old radio pulsars were 
 too faint for a detailed examination of their X-ray emission (cf.~Sun et 
 al.~1993; Manning \& Willmore 1994; Becker \& Tr\"umper 1997; Saito 1998).
 However, especially old rotation-powered non-recycled pulsars are of
 particular interest for the study of particle acceleration and high
 energy radiation processes near the neutron star's surface and in its
 magnetosphere. This is because their ages are intermediate between
 those of the well-studied young and cooling neutron stars, whose surface
 may produce copious thermal X-ray photons, and those of very old
 recycled millisecond pulsars, in which thermal hot-spot and
 non-thermal magnetospheric X-ray production mechanisms are believed
 to dominate.
 Old, non-recycled pulsars (outside globular clusters) therefore aid
 in answering questions such as how do the emission properties of the 
 younger pulsars, like Geminga, PSR B0656+14 and PSR B1055-52, change 
 as they age from $\sim 10^5$ to $10^7$ years? Will the thermal emission 
 simply fade away due to cooling with increasing age or will the star be 
 kept hot (at about $0.5-1 \times 10^5$ K) over millions of years due to 
 energy dissipation by processes such as internal frictional heating
 ($\dot{E}_{diss} \sim 10^{28}-10^{30}$ erg/s) and crust cracking, as
 proposed by vortex creeping and pinning models? What happens to the
 non-thermal, hard-tail emission seen in the X-ray spectra of the
 middle-aged field pulsars?  (See e.g.~Saito 1988; Becker \&
 Aschenbach 2002; De Luca et al.~2005). Will this emission become the
 dominant source or will this component also decay with time and will
 only thermal emission from the hot and heated polar-caps remain?

 In order to address these questions, we initiated a program to study
 the X-ray emission properties of old rotation-powered pulsars with
 XMM-Newton, aiming to probe and identify the origin of their X-ray
 radiation.  First results from parts of this project have been
 presented recently, reporting on the pulsars B0950+08, B0823+26,
 J2043+2740 (Becker et al.~2004; Zavlin \& Pavlov 2004) and PSR B0628-28 
 (Becker et al.~2005) all of which have a spin-down age in the range 
 from about one million to seventeen million years.

 If one extrapolates the X-ray emission properties of young and
 cooling neutron stars to this age bracket, one may expect that the
 cooling emission fades away and thermal emission from heated polar 
 caps dominates the X-rays. Surprisingly, the X-ray emission from 
 old pulsars is largely dominated by non-thermal radiation processes. 
 None of the pulsars' X-ray spectra required the addition of a 
 thermal component, consisting of hot polar cap emission, to model 
 their energy spectra. Further support for an emission scenario 
 dominated by non-thermal mechanisms is given by the 
 observed temporal emission properties. The pulse profiles of 
 PSRs B0950+08 and B0628-28 are not broad and sinusoidal as would have 
 been expected for spin-modulated thermal X-ray emission from heated 
 polar caps, but are double peaked with narrow pulse components and 
 pulsed fractions in the range of $\sim 30-50\%$.

 Some models, such as those by Harding \& Muslimov (2001; 2002; 2003),
 predicted in the framework of their revised space-charge-limited flow
 model that polar cap heating, as a fraction of the spin-down
 luminosity, increases with pulsar age and should be most efficient
 for pulsars of spin-down age {\Large $\tau$}$\sim 10^7$ yrs, if they
 are in fact producing pairs from curvature radiation
 photons. However, according to the Harding \& Muslimov model,
 B0950+08 and B0823+26 cannot produce pairs from curvature radiation
 of primary electrons since they both lie below the curvature
 radiation pair death line in the $P$-$\dot{P}$ diagram of radio
 pulsars (cf.~Becker et al.~2004 for a more detailed discussion).

 A good candidate object to test these models is \PSRB\, which according
 to its X-ray emission properties can be considered to be prototypical of 
 an old pulsar. With a pulse period of $P=226.5$~ms and a period
 derivative of $\dot{P}=1.16\times 10^{-15}$, its characteristic age
 is determined to be $\sim 3\times 10^6$ years. These spin parameters
 imply a spin-down luminosity of $\dot{E} = 3.9 \times 10^{33}\, \mbox{erg
 s}^{-1}$ and a magnetic field at the neutron star magnetic poles of\,
 $B_\perp \sim 5 \times 10^{11}$~G.  With a radio dispersion measure
 of $\,3.176\,\mbox{pc cm}^{-3}$, the NE2001 Galactic free electron
 density model of Cordes \& Lazio (2002) predicts a distance of 170
 pc.  However, the  recent astrometric measurements by
 Chatterjee et al.~(2004) yielded a precise proper motion and parallax
 determination that translates into an accurate distance measurement of
 $d=361^{+10}_{-8}$ pc and a transverse speed of
 $V_\perp=177^{+4}_{-5}$ km s$^{-1}$.

 Thus, \PSRB\ is among the closest pulsars known. In addition to its
 relatively young age it appears to be the brightest among all old
 non-recycled X-ray detected rotation-powered pulsars. Its X-ray
 emission was discovered with the EINSTEIN observatory by Helfand
 (1983). Pulsed X-ray emission was discovered using a deep ROSAT
 observation (Yancopoulos et al. 1994). The pulse profile was found to
 be very broad with a single pulse stretching across the entire phase
 cycle, markedly different from the sharp peak observed in the radio
 band. The fraction of pulsed photons in the $0.1-2.4$ keV band was
 determined to be $\sim 30\%$.
%
%
 The ROSAT data were not able to constrain the nature of the pulsar
 emission as a blackbody spectrum (representing thermal polar-cap
 emission) and a power law model (representing non-thermal magnetospheric
 emission) fitted the data equally well (Becker \& Tr\"umper 1997).
 Similar results were obtained in the analysis of ASCA spectral/timing 
 data by Wang and Halpern (1997) and Saito (1998).  The broad sinusoidal 
 pulse profile together with the higher column absorption of 
 $\sim (0.6-1.1)\times 10^{21}\,\mbox{cm}^{-2}$ deduced from power law 
 fits were taken as indirect arguments and strong indicator for a 
 thermal polar-cap origin of the X-rays\footnote{Fitting a blackbody 
 model resulted in a column density of $\sim (1-3) \times 10^{20}\,
 \mbox{cm}^{-2}$ which was believed to be in better agreement with the 
 close pulsar distance of $\sim 170$ pc as believed at that time.}.

 In contrast, Slowikowska et al.~(2005) recently found that a single
 blackbody spectral model cannot describe the pulsar spectrum if the
 ROSAT and ASCA observed spectra are modeled in a joint analysis.  In
 their work it is shown that a single power law model or a composite
 model consisting of a two component blackbody spectrum can successfully 
 describe the energy spectrum up to $\sim 7$ keV. The higher column 
 density fitted by these models is found to be in agreement with that 
 observed for other sources located near to the pulsar's line of sight 
 and at comparable distances (Slowikowska et al.~2005).

 Regardless of the nature of its X-ray emission, \PSRB\, seems to be
 special for its extended X-ray emission which was discovered by Wang,
 Li \& Begelman (1993) in their archival study of the Galactic soft
 X-ray background using deep ROSAT PSPC images. They found that the
 orientation of the diffuse X-ray emission is almost aligned with the
 pulsar's proper motion direction, suggesting an interpretation in
 terms of an X-ray emitting trail behind the pulsar. If indeed
 associated with the pulsar, the trail could account for $\sim 3
 \times 10^{-4}$ of the pulsar's spin-down luminosity although the
 effective brightness may depend strongly on the density of the
 ambient interstellar matter. In recent years, the lack of
 confirmation of the trail from a subsequent $\sim 350$ ksec deep
 ROSAT HRI observation casts some doubt on its existence.

 Near-UV emission from \PSRB\, has been detected in three partly
 overlapping spectral bands using the Hubble Space Telescope's Faint
 Object Camera (Pavlov et al.~1996) and the NUV-MAMA detector (Mignani
 et al.~2002). The nature of the optical emission is uncertain since 
 the paucity of color information makes any spectral fit based on the 
 optical data only merely tentative.

\begin{figure}
\centerline{\psfig{figure=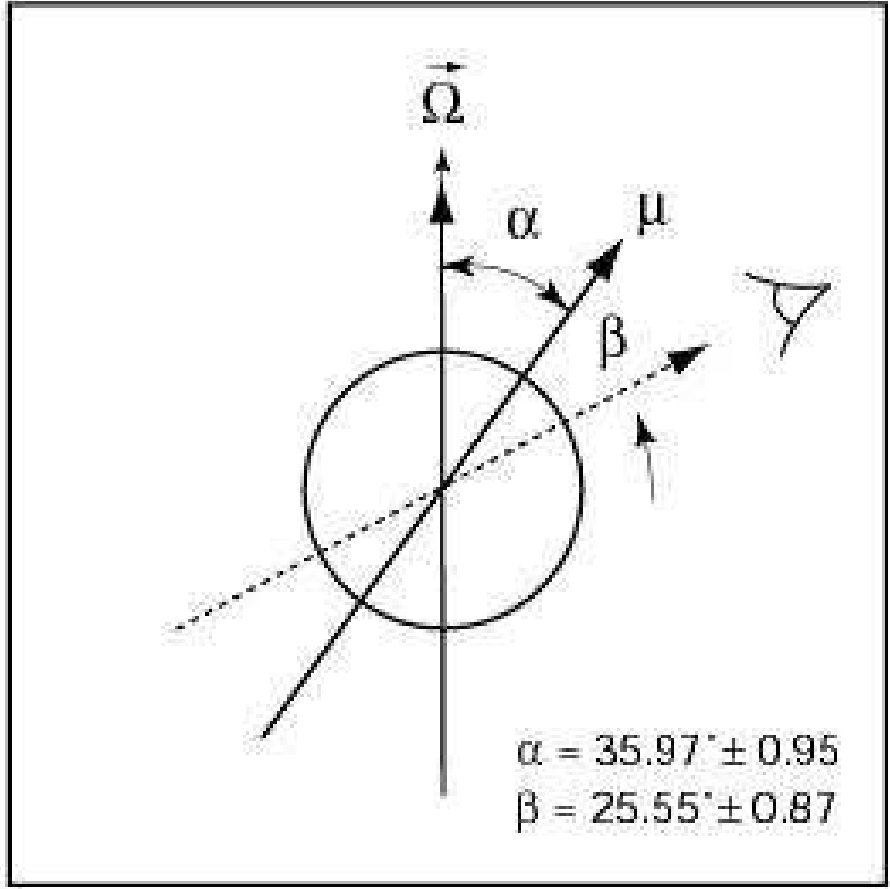,width=12cm,clip=}}
\caption[]{Emission beam geometry of \PSRB\, as deduced from fitting the
rotating vector model to the polarization angle swing observed at radio
frequencies (e.g.~Everett \& Weisberg 2001). $\alpha$ is the inclination of
the magnetic axis, $\beta$ the minimum angle between the magnetic axis and
the line of sight. $\vec{\Omega}$ is the rotation axis.} \label{PSRB_geometry}
\end{figure}

 \PSRB\, has also been extensively observed at radio frequencies. The 
 main peak of the radio profile, although much smaller than the X-ray 
 pulse, has a substructure which can be modeled by six separate 
 components (Kramer et al.~1994). Low-level emission connects this 
 main pulse with an
 interpulse about 180 deg in longitude apart (see e.g.~Everett \&
 Weisberg 2001).  The pulsar's viewing geometry has been studied by
 many authors via the observed polarization angle swing, applying a
 rotating vector model. Everett \& Weisberg (2001) reviewed the
 various results and concluded that both, the main and interpulse, are
 produced by nearly aligned rotation and magnetic axes and are emitted
 from nearly opposite sides of a wide, hollow cone. They derive an
 inclination of the magnetic axis with the spin axis of $\alpha\sim 36
 ^o$ while the impact angle of the line of sight was determined to be
 $\beta\sim 26^o$ (cf.~Fig.\ref{PSRB_geometry}).

 In this paper we report on X-ray, optical H-alpha and radio
 observations of \PSRB\, which were made with XMM-Newton, the
 ESO New Technology Telescope (NTT) in La Silla (Chile) and the
 Jodrell Bank Radio Observatory in order to explore the spectral
 and timing emission properties of this interesting pulsar and its
 environment. The paper is organized in the following manner: in \S2
 we describe the radio, optical H-alpha and XMM-Newton observations
 of \PSRB\, and its X-ray trail and provide the details of the data
 processing and data filtering. The results of the spatial, spectral
 and timing analysis are given in \S3. A summary and concluding
 discussion is presented in \S4.

\section{OBSERVATIONS AND DATA REDUCTION\label{obs}}

\subsection{RADIO OBSERVATIONS OF \PSRB\label{radio_obs}}

 The ephemerides for the analysis of the X-ray data were obtained from
 radio observations and the measurement of pulse times--of--arrival
 (TOAs) using the 76-m Lovell radio telescope at Jodrell Bank
 Observatory. Table \ref{t:radio} summarizes the radio ephemerides of
 \PSRB. A dual-channel cryogenic receiver system sensitive to two
 orthogonal polarizations was used predominantly at frequencies close
 to 1400 MHz. The signals of each polarization were mixed to an
 intermediate frequency, fed through a multichannel filter-bank and
 digitized. The data were de-dispersed in hardware and folded on--line
 according to the pulsar's dispersion measure and topocentric period.
 The folded pulse profiles were stored for subsequent analysis. In a
 later off--line processing step, any sub-integrations corrupted by RFI
 were removed, the polarizations combined and the remaining
 sub-integrations averaged to produce a single total--intensity profile
 for the observation.  TOAs were subsequently determined by convolving,
 in the time domain, the averaged profile with a template corresponding
 to the observing frequency. The uncertainty on the TOA was found using
 the method, described by Downs \& Reichley (1983) which incorporates
 the off--pulse RMS noise and the `sharpness' of the template.  These
 TOAs were transferred to an arrival time at the solar system
 barycenter using the Jet Propulsion Laboratory DE200 solar system
 ephemeris (Standish 1982). More details can be found in Hobbs et
 al.~(2004). Spectral data from \PSRB\, were obtained from the
 compilation of Maron et al.~(2000).

\subsection{OPTICAL OBSERVATIONS OF \PSRB \label{optical_obs}}

 We have performed optical H$_\alpha$ and R-band observations of the
 $4.25 \times 5.41$ arcmin sky region around \PSRB\, using the
 SUperb-Seeing Imager (SUSI2) at the focus of the ESO New Technology
 Telescope\footnote{See
 http://www.ls.eso.org/lasilla/sciops/ntt/index.html for a description
 of the ESO NTT and its instrumentation.} in La Silla (Chile).

 19 exposures of 600 s each ($\sim$3.2 hours total integration time)
 have been taken on July, 18 and 20 2004 through the H$_\alpha$ filter
 (central wavelength $\lambda=6555.28\,\AA$; $\Delta \lambda= 69.76\,
 \AA$). Additional 15 shorter exposures of $30-60$ s in the R-band
 filter (central wavelength $\lambda= 6415.8\, \AA$; $\Delta \lambda=
 1588.9\,\AA$) were taken in order to discriminate H$_\alpha$ line
 emission from the continuum.  In order to compensate for the $8''$
 gap between the two SUSI2 CCD chips, the exposure sequences were
 taken with a jitter pattern with typical offset steps of $20''$ in
 RA. All observations have been performed with average airmass of
 1.35, clear sky conditions and good seeing ($\sim 0.6''$).  Single
 exposures were corrected for the instrumental effects (bias and dark
 subtraction, flat-fielding), cleaned of bad columns and of cosmic
 rays hits through median filter combination. To account for the
 uneven exposure map due to the jitter pattern, both the final
 H$_\alpha$ and R-band images were exposure-corrected.

 The image astrometry was recomputed using as a reference the
 positions of a number of stars selected from the Guide Star Catalogue
 II (GSC-II), which has an intrinsic absolute astrometric accuracy of
 $\sim 0\farcs35$ per
 coordinate\footnote{http://www-gsss.stsci.edu/gsc/gsc2/GSC2home.htm}. The
 pixel coordinates of the reference stars have been computed by a
 two-dimensional Gaussian fitting procedure, and transformation from
 pixel to sky coordinates was then computed using the programme
 ASTROM\footnote{http://star-www.rl.ac.uk/Software/software.htm},
 yielding an rms of $\sim$ 0\farcs08 in both Right Ascension and
 Declination, which we assume representative of the accuracy of our
 astrometric solution.

\subsection{XMM-NEWTON OBSERVATIONS OF \PSRB \label{xray_obs}}

 \PSRB\, was observed by XMM-Newton\footnote{For a description of 
 XMM-Newton, its instrumentation and the various detector modes 
 available for observations see http://xmm.vilspa.esa.es/.} as part 
 of the European Photon Imaging Camera (EPIC)
 guaranteed time program. Observations were performed on 2003 November 10
 (XMM rev.~718) with a duration of $\sim 11$ ksec and five months
 later on 2004 April 27 (XMM rev.~803) and April 29 (XMM rev.~804) for
 a duration of $\sim 22$ ksec and $\sim 23$ ksec, respectively.  In
 all three observations the EPIC Positive-Negative charge depleted
 Silicon Semiconductor (PN camera) was used as the prime
 instrument. The two Metal Oxide Semiconductor cameras (MOS1 \& MOS2)
 were operated in PrimeFullWindow mode to obtain imaging and spectral
 data. The EPIC-PN camera was set up to operate in SmallWindow
 read-out mode which provides imaging, spectral and timing information
 with a temporal resolution of 5.67 ms which is more than sufficient 
 to resolve the 226 ms period of \PSRB. The SmallWindow mode was preferred 
 over other EPIC-PN imaging modes because of its higher temporal resolution, 
 albeit its $\sim 30\%$ higher dead-time caused a decrease in the net 
 exposure by $\sim 1/3$. The medium filter was used for the EPIC-PN and 
 MOS1/2 cameras in all observations to block optical stray light. Given
 the target flux, both the RGS and optical monitor are of limited use.
 A summary of exposure times, instrument modes and filters used for 
 the X-ray observations of \PSRB\, is given in Table ~\ref{t:xray_obs}.

 XMM-Newton data have been known to show timing discontinuities in the
 photon arrival times with positive and negative jumps of the order of
 one to several seconds (Becker \& Aschenbach 2002; Kirsch et
 al.~2004). While an inspection of the data processing log-files 
 reveals that none of the EPIC-PN data taken from \PSRB\, exhibit 
 such discontinuities, we nevertheless used a release track Version 
 of the XMM-Newton Standard Analysis Software (SAS) version 6.5 
 (released in August 2005) to process the EPIC-PN data. This software 
 detects and corrects most of the timing discontinuities during data 
 processing. In addition, known timing offsets due to ground station 
 and spacecraft clock propagation delays are corrected by this software 
 in using newly reconstructed time correlation (TCX) data 
 (cf.~Becker et al.~2006). Barycenter correction of the EPIC-PN data 
 and all other analysis steps were performed by using SAS Version 6.1. 
 Data screening for times of high sky background was done by inspecting 
 the light-curves of the EPIC-MOS1/2 and PN data at energies above 10 keV. 
 Apart from having a rather high sky background contribution in both 
 April observations, very strong X-ray emission from soft proton flares 
 is covering about half of these April data sets. The data quality of 
 the shorter November 2003 observation is not reduced by these effects.

 Using light-curves with 100 s bins, we rejected time intervals where the
 MOS1/2 had more than 130, 140 and 175 cts/bin in the 2003 November,
 2004 April 27 and 29 observations, respectively. For the EPIC-PN data 
 sets, we rejected times with more than 7, 12 and 40 cts/bin. The data 
 screening reduced the effective exposure time for the MOS1/2 and 
 PN-camera to a total of 64 ksec and 23.8 ksec, respectively.

 For the spectral analysis based on the MOS1/2 data we used only
 those events with a detection {\em pattern} between $0-12$
 (i.e. single, double and triple events) and the {\em flag} parameter
 set to less than, or equal to, 1. The latter criterion excludes events
 which are located near a hot pixel, or a bright CCD column, or
 which are near the edge of the CCD. For the EPIC-PN timing and
 spectral analysis, we used only single and double events, i.e.~those
 which have a {\em pattern} parameter of less than, or equal to, 4 and a
 flag parameter equal to zero. The energy range of the MOS1/2 and EPIC-PN 
 CCDs was restricted to $0.3-10$ keV for the spectral analysis due to 
 calibration issues towards softer spectral channels and to $0.2-10$ keV  
 for the timing analysis.

 \section{ANALYSIS OF THE MULTI-WAVELENGTH DATA OF \PSRB\label{PSRB}}

\begin{figure}
\centerline{\psfig{figure=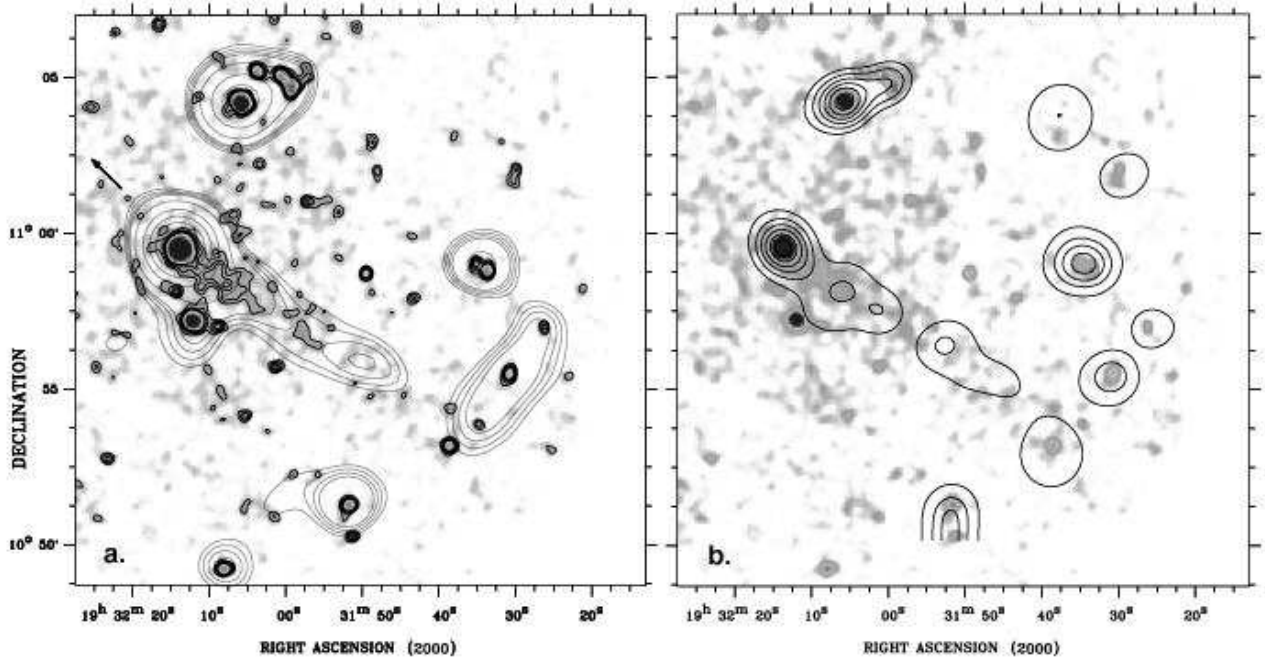,width=17cm,clip=}}
\caption[]{{\bf a.}\, \PSRB\, and its environment as seen in the 
$0.2-10\,\mbox{keV}$ energy band by XMM-Newton's MOS1/2 cameras. The image 
has been smoothed with a $3\sigma$ Gaussian filter. Contour lines in the 
MOS1/2 image are at levels of $(2.0 - 3.2) \times 10^{-2}$ EPIC-MOS1/2
counts/arcsec$^2$.  Contour lines from the ROSAT PSPC image at levels of
$(0.98 - 3.11) \times 10^{-2}$ PSPC counts/arcsec$^2$ are overlaid 
with a 50\% reduced opacity for comparison. The pulsar's proper motion 
direction is indicated by an arrow. {\bf b.}\, MOS1/2 image overlaid 
with ROSAT HRI contour lines at levels of $(2.86 - 9.1) \times 10^{-1}$
HRI counts/arcsec$^2$.} \label{PSRB_XMM_field_vs_PSPC_HRI_field}
\end{figure}

 The X-ray counterpart of \PSRB\, is detected with high significance
 in both the MOS1/2 and EPIC-PN data. The count rates are $0.0283
 \pm 0.0001$ cts/s (MOS1/2) and $0.078 \pm 0.003$ cts/s (EPIC-PN)
 within the $0.2-10$ keV band. Inspection of the MOS1/2 and EPIC-PN
 images revealed diffuse extended emission at the position of the
 putative X-ray trail seen in the ROSAT data by Wang, Li and Begelman
 (1993). Figure \ref{PSRB_XMM_field_vs_PSPC_HRI_field} shows the 
 MOS1/2 image made with the 2003 November and 2004 April data. Contour 
 lines indicate the diffuse X-ray trail lying in the direction opposite 
 to the transverse motion of the pulsar ($\mu_{\rm RA} = 17.00\pm0.27$ 
 mas yr$^{-1}$ and $\mu_{\rm DEC} =-9.48\pm 0.37$ mas yr$^{-1}$, 
 Chaterjee et al.~2004). Contour lines, obtained from a re-analysis of 
 45 ksec ROSAT PSPC and 350 ksec ROSAT HRI observations, are overlaid. 
 Owing to a sensitivity ratio between the PSPC and HRI of up to a factor of 
 $\sim 5$ (depending on the source spectrum) and a much higher detector
 noise the faint trail like emission is much harder to detect in the HRI
 data. Only after applying an adaptive kernel smoothing 
 procedure a source structure which matches with the shape of the trail 
 seen in the PSPC and MOS1/2 images becomes visible in this data. An
 interesting difference, though, is that in the HRI data the trail seems 
 to break up into two separate pieces. In the MOS1/2 images this region 
 is partly covered
 by a CCD gap. The angular resolution of the ROSAT detectors are 25 arcsec 
 in the PSPC and 5 arcsec in the HRI, respectively, while that of XMM-Newton
 is 15 arcsec (Half Energy Width). 

\begin{figure}
\centerline{\psfig{figure=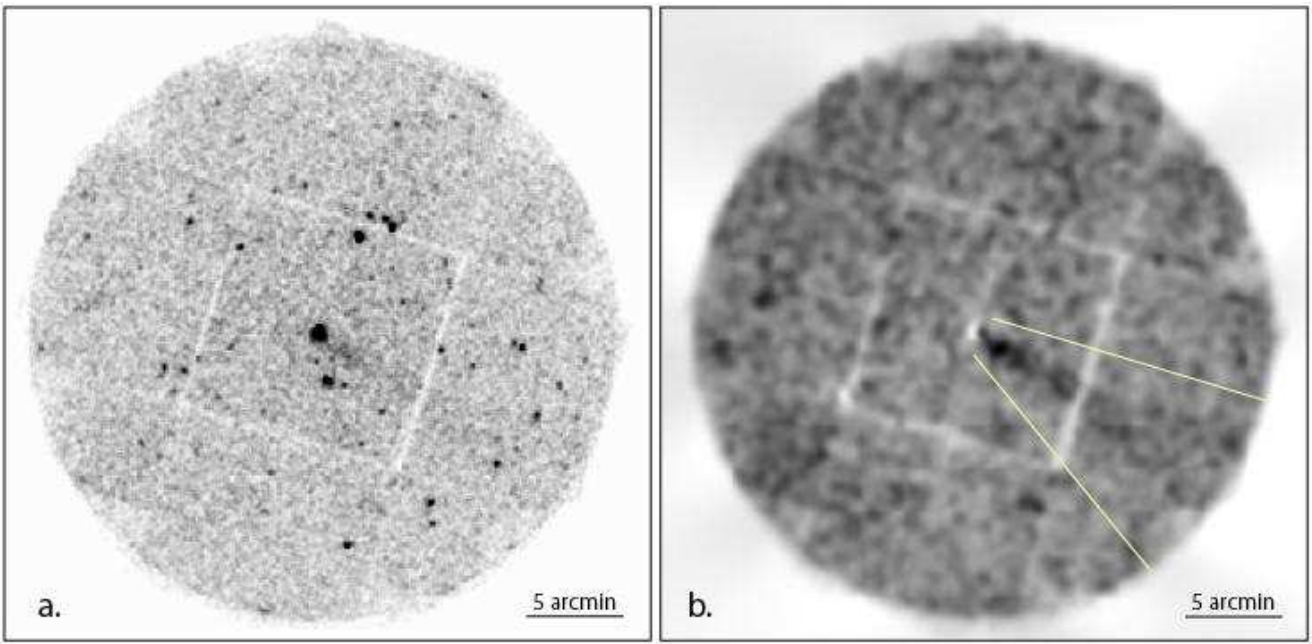,width=17.5cm,clip=}}
\caption[]{{\bf a.}\, XMM-Newton MOS1/2 full 15 arcmin field of view of the
sky region around \PSRB\, (central source) as seen in the $0.2-10$ keV energy 
band. {\bf b.}\, Same image than shown in the left panel but with point sources 
removed and vignetting correction applied. The image has been smoothed with 
a $3\sigma$ Gaussian filter. The segment of interest is indicated.} 
\label{PSRB_mos_fullfield} 
\end{figure}

 Inspecting the MOS1/2 full field of view (cf.~Figure \ref{PSRB_mos_fullfield}a)
 there is some indication for trail emission beyond the edge of the inner MOS CCD. 
 To reduce the impact on the trail detection due to the presence of other X-ray 
 point sources in the field (though no bright sources are located along the trail) 
 we have removed point source contributions from the MOS1/2 data and corrected 
 the image for telescope vignetting effects. The result is displayed in 
 Figure \ref{PSRB_mos_fullfield}b. Clearly, there is a whiff of emission which 
 probably extends down to the edge of the detector's field of view, but its 
 significance at locations more distant from the pulsar is gradually fading into 
 the background.  In order to estimate the significance of the emission along the 
 tail we extracted four circles at increased distance from the pulsar, at 1.5', 
 4.5', 7.5', and 10.5', respectively. The significance of the trail emission in 
 these circles is computed to be  
 (C1): $493\,\mbox{scts}/\sqrt{1417\,\mbox{bcts}} = 13\,\sigma$, 
 (C2): $183\,\mbox{scts}/\sqrt{1197\, \mbox{bcts}}= 5.3\,\sigma$, 
 (C3): $161\,\mbox{scts}/\sqrt{1234\, \mbox{bcts}} = 4.6\,\sigma$, and (C4):
 $125\,\mbox{scts}/\sqrt{1100\, \mbox{bcts}} = 3.8\,\sigma$, where  {\em scts} and 
 {\em bcts} denote the source counts and background counts measured for these
 four circles in the energy band $0.2-10$ keV.

\begin{figure}
\centerline{\psfig{figure=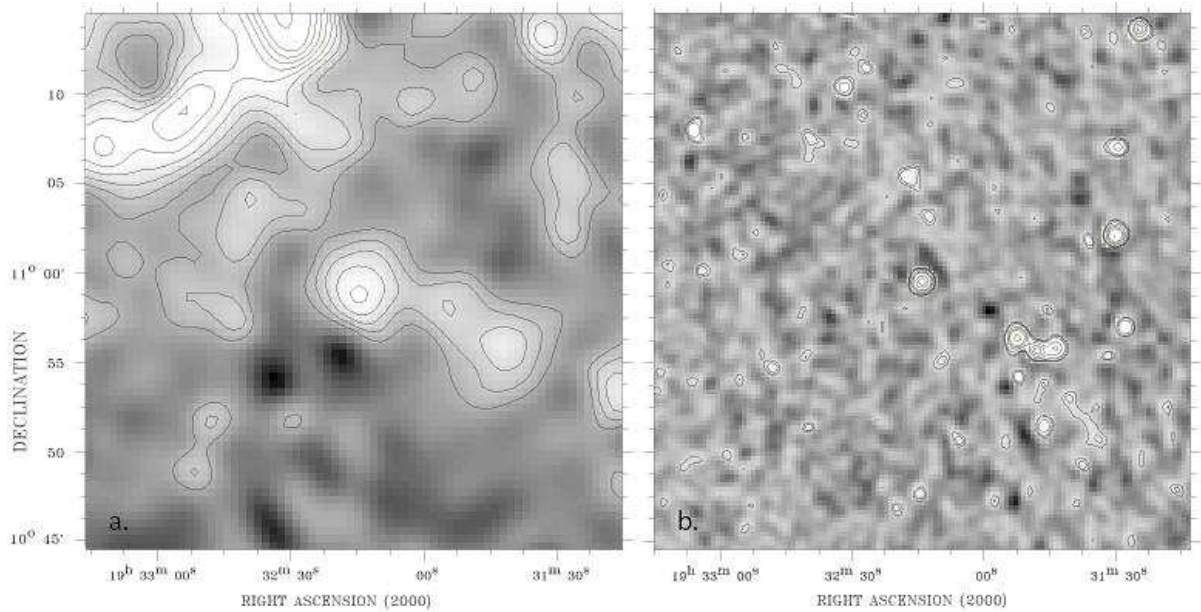,width=16cm,clip=}}
\caption[]{{\bf a.}\, A $30 \times 30$ arcmin section of the 11cm (2.72 GHz)
 radio map around \PSRB\, as seen with the Effelsberg 100m Radio Telescope. 
 The map is centered on the pulsar. The smooth galactic background has
 been removed and the image was enhanced by unsharp masking on a scale
 of 7.5 arcmin. Beam size (HPBW) was 4.3 arcmin and the rms 7 mJy. The 
 Pulsar itself is seen as a polarized source with a flux of $30-40$ mJy. 
 Polarization of the other features was not detectable at the sensitivity 
 of the survey. {\bf b.}\, Same sky region than shown in the left panel but
 observed during the NRAO VLA Sky Survey at 1.4 GHz with a beam size of 45 
 arcsec (FWHM).} \label{Effelsberg_NVSS_map}
\end{figure}

 In the first half of the 80s the Effelsberg 100-m radio telescope has
 mapped the entire Galactic Plane in the latitude range $\pm 5^\circ$
 at 11 cm wavelength (Reich et al.~1990). Inspecting this data for a
 possible radio counterpart of the pulsar's X-ray trail we discovered
 an elongated feature which roughly matches with that observed in the 
 X-ray band. Figure \ref{Effelsberg_NVSS_map}a shows the $\sim 30 \times 
 30$ arcmin region centered on \PSRB\, as observed in the 11 cm Effelsberg 
 survey. The extracted map has 140 x 140 pixels. Its resolution (HPBW) is 4.3'
 with fluxes ranging from 0 to 250 mKTb (1 mKTb = 0.398 mJy). The rms of
 the survey was given as 20 mKTb (= 8 mJy). In the vicinity of  the source 
 we found a diffuse background of 25 mJy. The pulsar itself was evident with
  25 mJy above the local background level. The radio tail was visible with 
 fluxes of up to 15 mJy above local background and unresolved in the transverse
 direction. The tail was not aligned with the low level residual scanning effects.
 The distance to the bright node on the tail was found to be 8.8', but more aligned
 features were visible further in the beam out to 12'. In order to estimate the
 probability of a spurious detection we derived a probability of a pixel 
 exceeding 34 mJy as $p_x(\rm {34mJy})=0.374$ from a histogram of all pixels in our map.
 We made another histogram from a 33' (78 pixel) line along the tail direction from the pulsar, excluding
 the first 6' of the point source. 18 of the remaining 64 pixels were found to be
 above 34 mJ. The binominal distribution gives a probability of $p_{rand}=0.032$
 for a random occurrence of such a result along an arbitrary line of similar length in our map.
 Using the resolution (4.3') for a conservative estimate of the tail width, we find that
 the probability of a random alignment in the direction of proper motion is about
 $p_{align}=0.078$. Hence the probability of a coincidence of  weak random radio
 sources with the observed psr tail is about $p_{radio}= p_{align} p_x = 2.5\cdot 10^{-3}$

\begin{figure}
\centerline{\psfig{figure=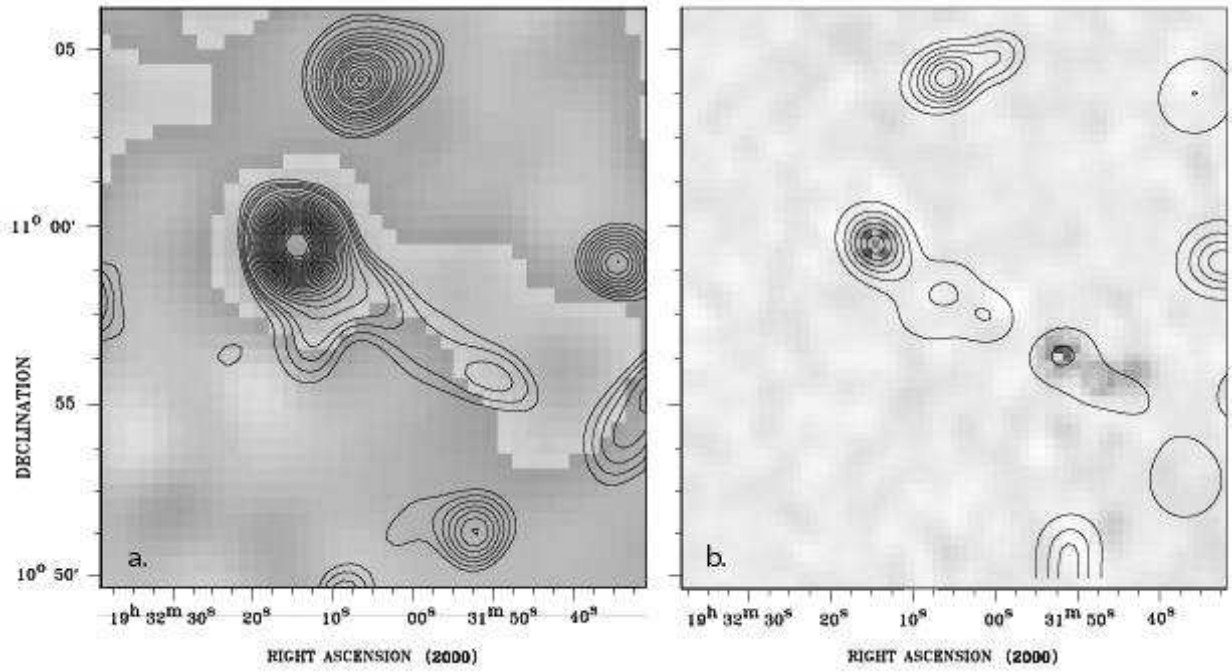,width=16.5cm,clip=}}
\caption[]{{\bf a.}\, Contour lines from the ROSAT PSPC overlaid on part of the 
 Effelsberg 11cm radio image. {\bf b.}\,  Contour lines from the ROSAT HRI overlaid 
 on part of the 1.4 GHz NRAO VLA Sky Survey image.} \label{PSRB_radio_vs_rosat}
\end{figure}

 Nevertheless, inspecting the NRAO VLA Sky Survey (NVSS) which is a 1.4 GHz
 continuum survey mapping the entire sky north of -40 deg declination (Condon et 
 al.~1998) we found faint not fully resolved radio sources which match in position 
 with the brighter end of the radio tail seen at 11cm (2.72 GHz). There is no 
 radio emission seen bridging the pulsar and these faint sources, however, this 
 absence is likely a function of the VLA telescope configuration which affects the 
 sensitivity of the observations to diffuse emission. The NVSS image
 has an angular resolution of 45 arcsec (FWHM) and is shown in Figure 
 \ref{Effelsberg_NVSS_map}b. An elongated emission feature seen at RA(2000) 19$^h$ 
 31$^m$ 36$^s$, DEC +10$^d$ 51' 54'', and thus still located within the segment 
 indicated in Figure \ref{PSRB_mos_fullfield}b, roughly lines up with the pulsar's 
 proper motion direction. Making use of such an alignment to claim a relation with 
 \PSRB, however, seems  premature. As there are other bright sources in the 11cm 
 Effelsberg data which are not seen in the NVSS image (e.g.~the bright circularly 
 shaped source located in the north-east of Figure \ref{Effelsberg_NVSS_map}a) the 
 NVSS data cannot help to constrain the nature of the elongated radio feature 
 seen at 11cm. A comparison of the Effelsberg and NVSS radio images with the ROSAT PSPC
 and HRI X-ray images is shown in Figure \ref{PSRB_radio_vs_rosat}. Interestingly, the 
 part of the tail which in the HRI appears to be separated from the pulsar matches with
 the position of the three not fully resolved radio sources in the NVSS, suggesting 
 that both could be related. 

\begin{figure}
\centerline{\psfig{figure=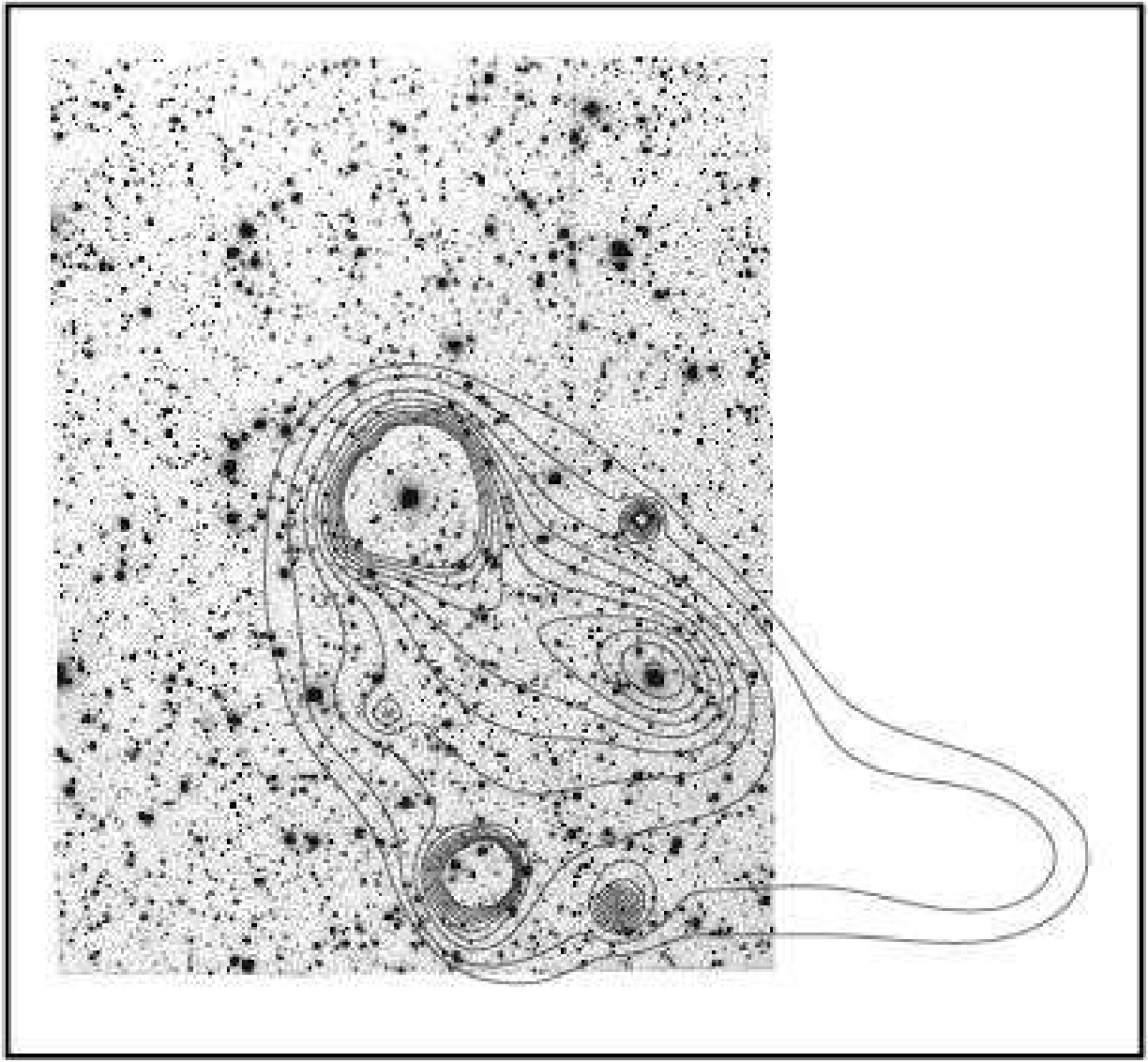,width=16.5cm,clip=}}
\caption[]{The $4.25\times 5.41$ arcmin sky field around \PSRB\ as seen in
in the light of H$_\alpha$ with the ESO NTT. Overlaid are the
contour lines from the XMM-Newton MOS1/2 image to which adaptive
kernel smoothing was applied.}
\label{PSRB_halpha}
\end{figure}

 The NTT H$_\alpha$ image which shows the $4.25\times 5.41$ arcmin sky region around 
 \PSRB\, is shown in Figure \ref{PSRB_halpha}. Contour lines from the MOS1/2 image are 
 overlaid. The inspection of the H$_\alpha$ image and of the ratio of the H$_\alpha$ to 
 $R$-band images does not reveal any diffuse structure that could be related to the 
 brighter parts of the pulsar's X-ray trail. We performed a rough flux calibration of 
 the H$_\alpha$ image by computing the H$_\alpha$ fluxes corresponding to the R magnitudes 
 of a sample of GSC-II stars which are included in the field assuming blackbody spectra with 
 the appropriate stellar temperatures. We then derived the relation between the instrumental 
 magnitude and the H$_\alpha$ fluxes. A flux upper limit of $\sim 10^{-16}\,\mbox{erg s}^{-1}
 \mbox{cm}^{-2}\mbox{arcsec}^{-2}$ in the H$_\alpha$ band can be set for the sky region 
 around the pulsar.

 As can be seen in Figure \ref{PSRB_halpha}, there is a bright star
 near to the position of \PSRB. A second bright star seems to coincide
 with the brighter part of the trail near to the pulsar. In view of
 the density of stars in the observed field the chance probability to
 have a star near to the position of an X-ray source is quite high. We
 investigated therefore whether these stars contribute to the flux
 recorded from \PSRB\, and the brighter part of its trail.  The bright
 star close to \PSRB\, has been identified to be of K(4-6)III-I type 
 (Kouwenhoven \& van der Berg 2001) and has a magnitude in B and R of 
 $\sim 14.85$ and $\sim
 13.44$, respectively. The bright star within the trail is possibly of
 K2-4 class and according to the GSC-II and the 2MASS catalogues has
 B, R, J and K magnitudes of $\sim 13.97, \sim 12.13, \sim 10.6$ and
 $\sim 10.0$, respectively.  With $\log(F_x/F_{opt}) = -2.77 \pm 1.0$
 (Krautter et al.~1999), we find for a mekal plasma model with
 kT=0.35, solar abundances, and a column absorption of $N_H\sim
 10^{21}\,\mbox{cm}^{-2}$ a possible contribution from these stars to
 the soft channels in XMM-Newton of $\le 10- 20\%$. Given the very
 large spread in the emission properties of K stars and taking its
 colors into account the star near to \PSRB\, is possibly a bright
 giant, and thus could lie towards the lower end of the $F_x/F_{opt}$
 range (e.g.~Zickgraf et al.~2005) which then would imply a negligible
 flux contribution in the pulsar extraction region. As far as the
 pulsar's X-ray trail is concerned it is clear from its length and
 from its hard X-ray spectrum (cf.~\S\ref{PSRB_spect_diff}) that this
 trail is not due to unresolved foreground or background sources.

 \subsection{TIMING ANALYSIS\label{PSRB_timing}}

 We used all the EPIC-PN SmallWindow mode data for the timing
 analysis, including those times of high sky background which were
 excluded for spatial and spectral analysis. Experience has shown that
 this does not affect the results from the timing analysis if the sky
 background is properly taken into account in all pulsed-fraction
 estimates.

 Events were selected from a circle of 20 arcsec radius centered on
 the pulsar radio timing position (cf.~Table 1).
 For the barycenter correction we applied the standard
 procedures for XMM-Newton data using {\em barycen-1.17.3}\/ and the
 JPL DE200 Earth ephemeris (Standish 1982) to convert photon arrival
 times from the spacecraft to the solar system barycenter (SSB) and
 the barycentric dynamical time (TDB). The pulsar radio timing
 position (cf.~Table \ref{t:radio}) was used for the barycenter
 correction. The spin-parameters $f$ and $\dot{f}$ of \PSRB\, are
 known with high precision from our contemporaneous radio
 observations, covering all XMM-Newton orbits relevant for our
 analysis. \PSRB\, is not known to show timing irregularities
 (glitches) so that we can fold the photon arrival times using the
 pulsar's radio ephemeris. The statistical significance for the
 presence of a periodic signal was obtained from a $Z^2_n$-test with
 $1-10$ harmonics in combination with the H-Test to determine the
 optimal number of harmonics (De Jager 1987; Buccheri \& De Jager
 1989). The optimal number of phase bins for the representation of the
 pulse profile was determined by taking into account the signal's
 Fourier power and the optimal number of harmonics deduced from the
 H-Test (see Becker \& Tr\"umper 1999 and references therein).

 Within the $0.2-10$ keV energy band, 5736 events were available for
 the timing analysis of which $\sim 51\%$ are estimated to be
 instrument and sky background.  The $Z^2_n$-test gave 82.14 for $n=6$
 harmonics ($Z^2_1=48.61$). According to the H-Test, the probability
 of measuring $Z^2_6=82.14$ by chance is $\sim 1.6 \times
 10^{-12}$. The significance of the pulsed signal thus is comparable
 with that found recently in the other old pulsars B0950+08 (Becker et al.~2004)
 and PSR B0628-28 (Becker et al.~2005).

\begin{figure}
\centerline{\psfig{figure=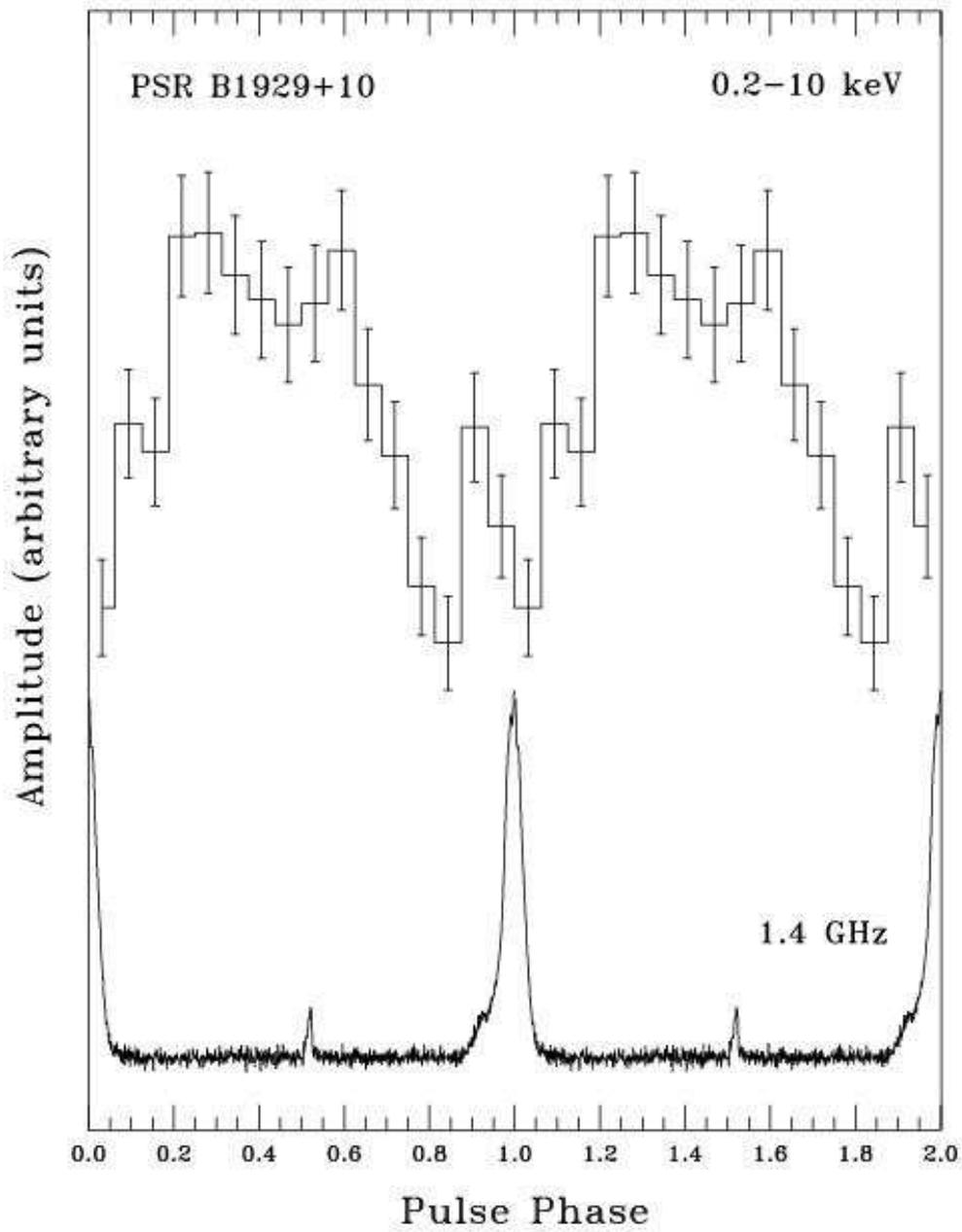,height=17cm,clip=}}
\caption[]{Integrated pulse profiles of \PSRB\, as observed in the $0.2-10$ keV
band with the EPIC-PN Camera (top) and at 1.4 GHz with the Jodrell Bank
radio observatory (bottom). The radio profile is plotted on a logarithmic scale
to better visualize the weak interpulse which has a displacement of $\sim 180^\circ$
from the main radio pulse peak. X-ray and radio profiles are phase related. Phase zero
corresponds to  MJD=53120.01448082337447 (TDB at SSB). Two phase cycles are shown
for clarity.} \label{PSRB_x_radio_profiles}
\end{figure}

\begin{figure}
\centerline{\psfig{figure=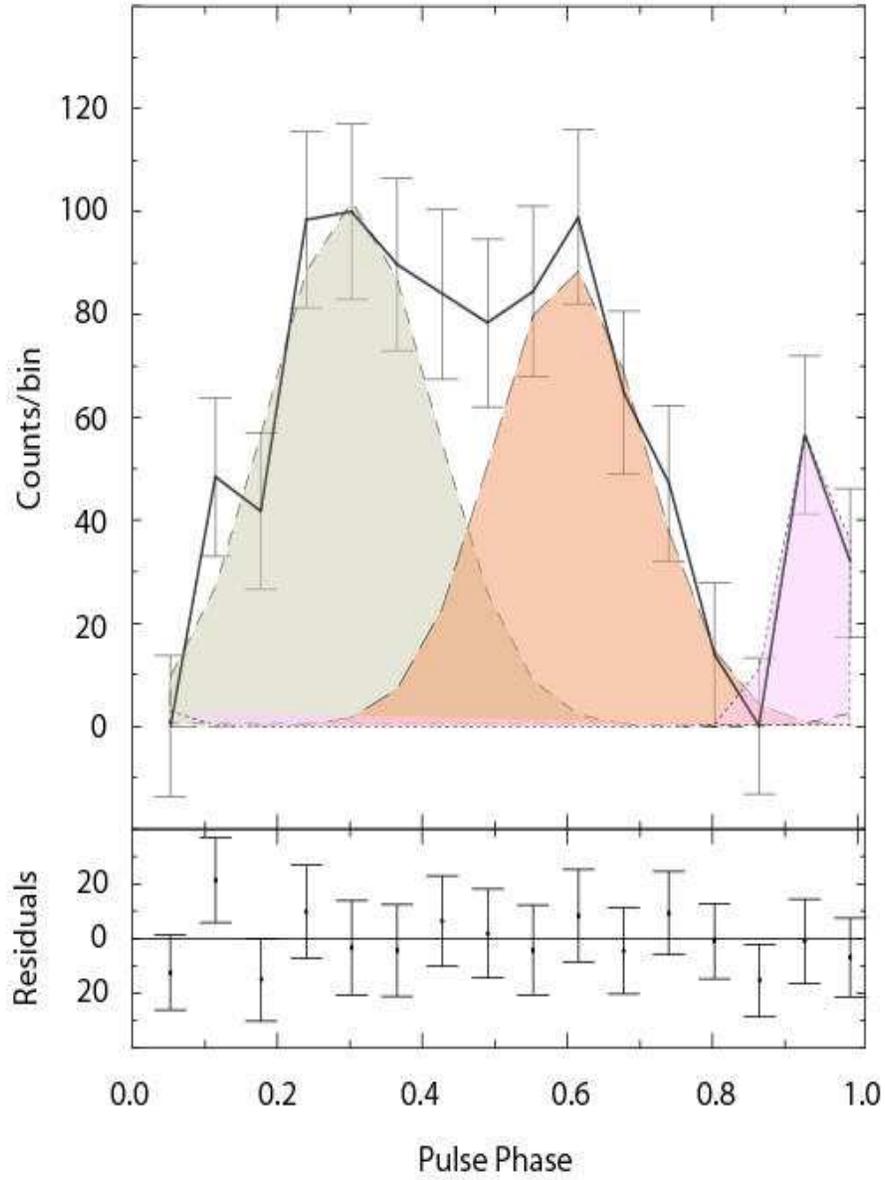,height=16cm,width=12cm, clip=}}
\caption[]{Results of fitting three Gaussians to the pulse profile of \PSRB\, observed in the
$0.2-10$ keV energy band.  The fit residuals in units of counts/bin are given in the lower panel.}
\label{PSRB_three_comp_fit}
\end{figure}

 The $0.2-10$ keV pulse profile is shown together with a radio profile
 observed at 1.4 GHz in Figure \ref{PSRB_x_radio_profiles}. It reveals a
 significant deviation from a sinusoidal pulse shape. The X-ray pulse
 profile is found to consist of at least two pulse peaks, a broader
 component of $\sim 273^\circ$ width, plus a narrow component of
 $\sim 44^\circ$ width. Taking the center of mass of the pulse as
 a reference point, the narrow X-ray pulse appears to be slightly
 phase shifted (by $\sim 22^\circ$) from the location of the main
 radio peak, although both components overlap well in phase. The
 broader pulse appears to have substructures hinting the presence of
 two  narrower pulse peaks which are not fully resolved by the 
 available data. Indeed, modeling the profile
 with three Gaussians yields acceptable results, supporting such an
 interpretation.  The fitted functions together with the post-fit
 residuals are shown in Figure \ref{PSRB_three_comp_fit}. The centers
 and widths (FWHM) of each of these three components  obtained from
 the fit are, in pulse longitudes, $(112^\circ, 80^\circ)$,
 $(220^\circ, 75^\circ)$, and $(346^\circ, 22^\circ)$, respectively.
 The uncertainty in the center position of each component is estimated
 as $\sim 8^\circ$ $(1\sigma$).

 In order to search for any energy dependence in the pulsar's temporal
 X-ray emission properties, we restricted the timing analysis to the
 $0.2-1.0$ keV, $1.0-2.1$ keV, and $2.1-10$ keV energy bands. This
 analysis shows that the pulsed signal appears to be most significant
 if we consider only events which are recorded at energies below $\sim
 2.1$ keV. The maximum $Z^2_n$-values found for the pulsed signal in
 the three energy bands are $Z^2_1=32.12$, $Z^2_3=52.1$, and
 $Z^2_6=28.4$, respectively. The fact that the pulsed signal is less
 significant beyond $\sim 2.1$ keV can be explained by an increase of
 instrumental background along with a decrease of pulsar signal, yielding
 a lower signal-to-noise ratio compared to the $0.2-1.0$
 keV and $1.0-2.1$ keV energy bands. An inspection of the pulse
 profiles for these two energy bands reveals that the emission from
 the narrow pulse peak appears only in the profile beyond $\sim 1$
 keV. Computing the fraction of pulsed photons (pulsed fractions,
 hereafter), we find $32 \pm 4 \%$
 for the total $0.2-10$ keV energy band.  The pulsed fractions in the
 sub-bands $0.2 - 1.0$ keV, $1.0-2.1$ keV, and $2.1-10$ keV are $24
 \pm 5 \%$, $44 \pm 6\%$, and $17 \pm 17 \%$, respectively (errors
 represent the $1\sigma$ confidence limits). For the narrow peak which
 appears at energies $\ge 1$ keV, the pulsed fraction in the $1.0-2.1$
 keV band is slightly higher than within the $0.2 - 1.0$ keV band,
 although the significance for this is at the $\sim 2\sigma$ level
 only.

 Computing the TOAs of the X-ray pulse we note that uncertainties of
 the XMM-Newton clock against UTCs are not relevant as those are on a
 scale of $\sim 100\,\mu s$ (Becker et al.~2005 in prep.) and thus are
 a factor of $\sim 140$ smaller than the bin width of the X-ray pulse
 profile shown in Figure \ref{PSRB_x_radio_profiles}. The largest
 uncertainty from comparing radio with X-ray pulse arrival is the
 definition of a suitable reference point which we choose to be the
 center of mass of a pulse peak.

 \subsection{SPECTRAL ANALYSIS\label{PSRB_spect}}

 \subsubsection{\PSRB}

 The energy spectrum of \PSRB\, was extracted from the MOS1/2 data by
 selecting all events detected in a circle of 40 arcsec radius,
 centered on the pulsar position. According to the XMM-Newton/EPIC-MOS 
 model point spread function, $\sim 87\%$ of all events of a point source
 are within this region. The background spectrum was extracted from a
 source-free circular region of 30 arcsec radius, northwest from the
 pulsar at RA(2000) $19^h\, 32^m\, 08.59^s$, DEC $10^\circ\, 59'\,
 36.49\arcsec$.  A second background spectrum was extracted from a
 circle of the same size but centered east from the pulsar at the
 position RA(2000) $19^h\, 32^m\, 19.5^s$, DEC $10^\circ\, 59'\,
 56.1\arcsec$. This second background spectrum was used to check the
 independence of the spectral results with respect to the selected
 background region.

 For the EPIC-PN data we used an extraction radius of 30 arcsec
 centered on \PSRB. This selection region includes $\sim 85\%$ of the
 point source flux. Out-of-time events and a gradient of decreasing
 background towards the PN-CCD readout node requires that the
 background spectrum be selected from regions located at about the
 same CCD row level as the location of the pulsar. We therefore
 extracted two background spectra from source-free circular regions of
 radius 30 arcsec about 1.5 arc-minute east and west of the pulsar
 position. The spectral results were found to be independent of the
 specific background region used.

 Because of its unique soft response we made use of archival ROSAT PSPC
 data and extracted the pulsar spectrum from a 60 arcsec circular region
 in the PSPC. The background spectrum was extracted from an annulus of
 70 arcsec and 100 arcsec inner and outer radius, respectively, centered
 on the pulsar.

 In total, the extracted spectra include 2405 EPIC-PN source counts
 and 1536 EPIC-MOS1/2 source counts. The PN and MOS1/2 spectral data
 were dynamically binned so as to have at least 30 counts per bin. 
 462 additional source counts, recorded within $0.1-2.4$ keV by ROSAT,
 were available for the joined ROSAT plus XMM-Newton spectral analysis
 of \PSRB. As the energy resolution of the ROSAT PSPC was only 
 $\sim 30\%$ at 1 keV (Briel et al.~1989) we binned the ROSAT data 
 so as to have 4 independent spectral bins. Owing to its unique soft 
 response, ROSAT data contribute information primarily to the lowest 
 spectral channels.
 
 Model spectra were then simultaneously fitted to the ROSAT and XMM-Newton 
 pulsar data. An anomaly in the MOS2 spectral data below 
 $\sim 0.8$ keV, which probably is related to the MOS redistribution 
 problem in which events from higher energies incorrectly are redistributed 
 downwards (cf.~Sembay et al.~2004), required exclusion of those MOS2 
 spectral bins from the analysis. 

\begin{figure}
\centerline{\psfig{figure=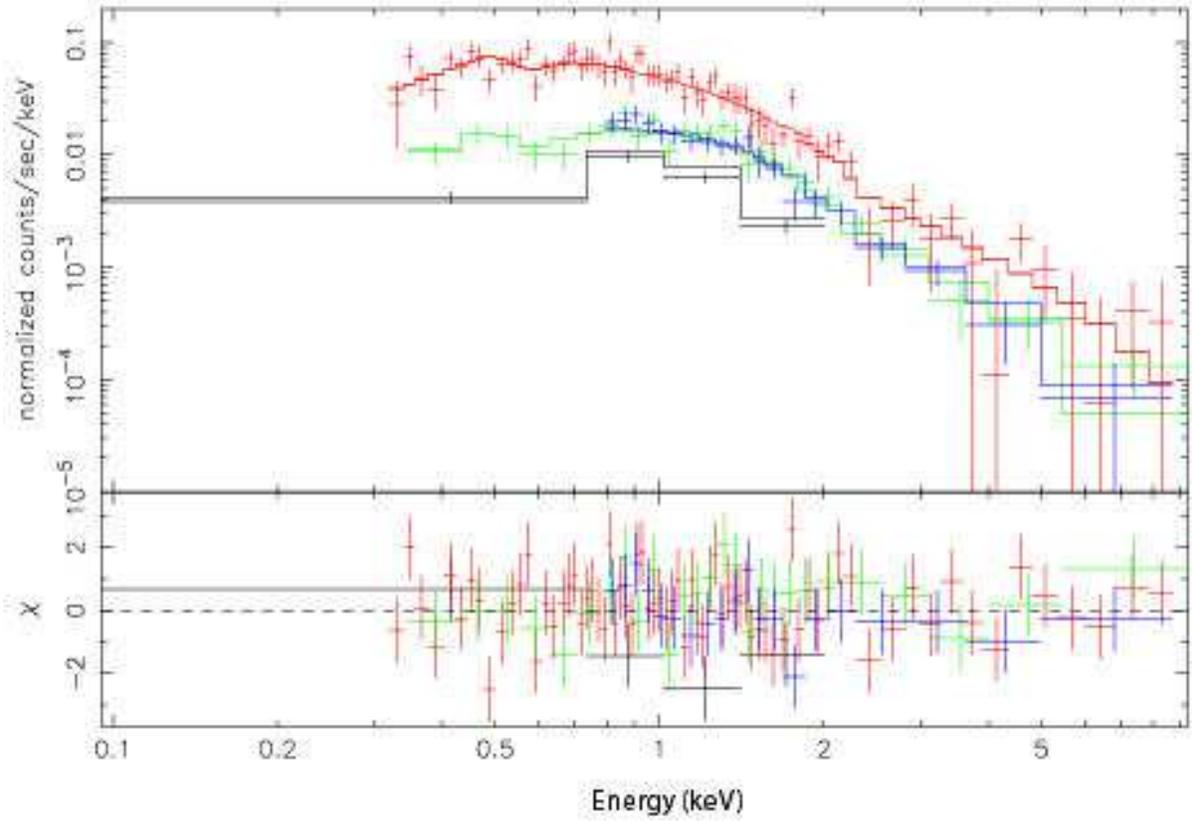,width=16cm,clip=}}
\caption[]{Energy spectrum of \PSRB\, as observed with the EPIC-PN (upper spectrum),
the MOS1/2 detectors (middle spectra), and the ROSAT PSPC (lower spectrum) and
simultaneously fitted to an absorbed power law model ({\it upper panel}) and
contribution to the \chisq\, fit statistic ({\it lower panel}).} \label{PSRB_pl_spectrum}
\end{figure}

\begin{figure}
\centerline{\psfig{figure=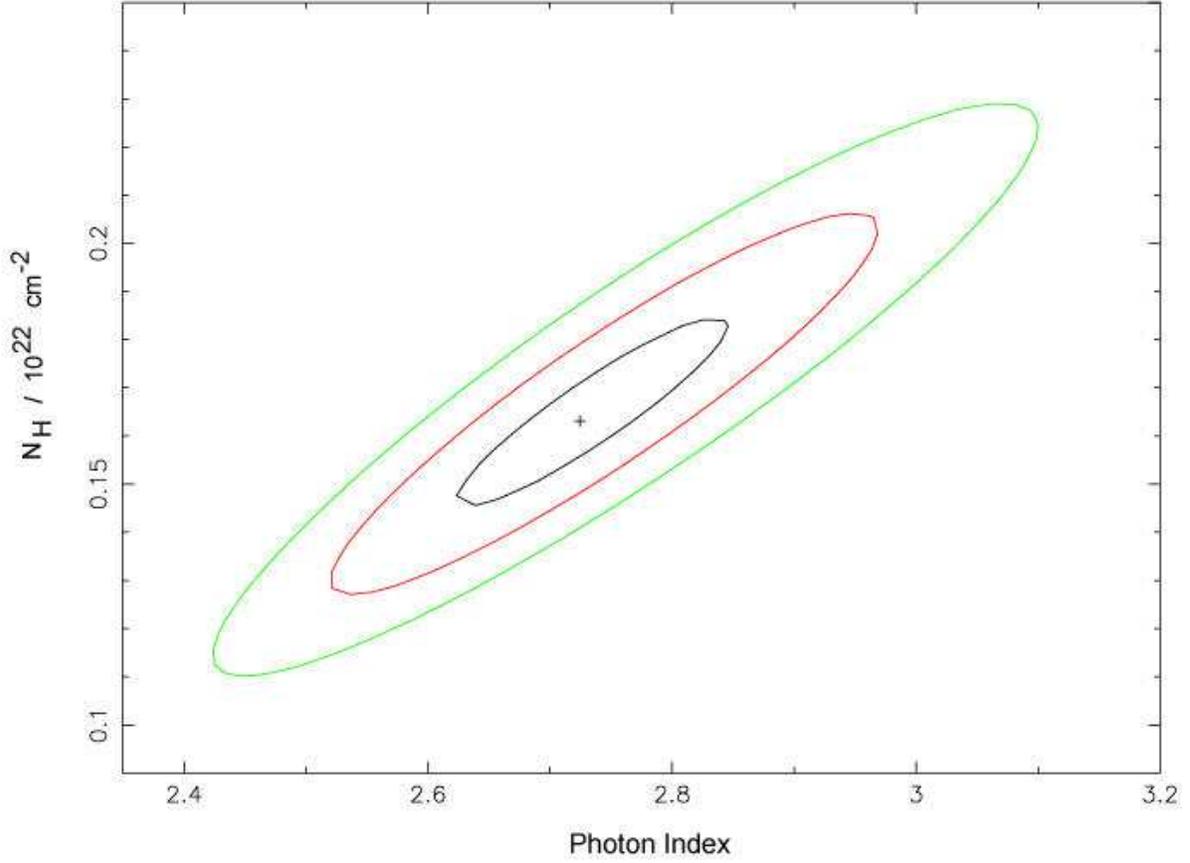,width=16cm,clip=}}
\caption[]{Contour plot showing the relative parameter dependence of the photon
index vs.~column absorption for the power law fit to the \PSRB\, XMM-Newton and ROSAT
data. The three contours represent the $1\sigma$, $2\sigma$ and $3\sigma$ confidence contours
for one parameter of interest. The `+' sign marks the best fit position,
corresponding to $\chi^2_{min} =0.989$ for 121 dof.}
\label{PSRB_pl_contour}
\end{figure}

 Amongst the single component spectral models, a power law model was
 found to give the statistically best representation (\chisq =119.6
 for 121 dof) of the observed energy spectrum. A single blackbody
 model (\chisq =246.9 for 121 dof) which was used by Yancopoulos et
 al.~(1994) and Wang \& Halpern (1997) to describe the ROSAT and ASCA
 observed pulsar spectrum did not give acceptable fits and is finally
 rejected. The best-fit power law spectrum and residuals are shown
 in Figure~\ref{PSRB_pl_spectrum}. Contour plots showing the
 relationship between the photon index and the column absorption for
 various confidence levels are shown in Figure \ref{PSRB_pl_contour}.

 In the following, we describe various models fitted to the energy
 spectrum of \PSRB. The resulting spectral parameters are summarized
 in Tables \ref{spectral_fits} and \ref{spectral_fits2} where all
 errors represent the $1\sigma$ confidence range computed for one
 parameter of interest.  The power law model yields spectral
 parameters which are in good agreement with the results obtained by
 Slowikowska et al.~(2005) based on their joint analysis of ROSAT plus
 ASCA data. The unabsorbed energy fluxes and luminosities (see
 Table~\ref{spectral_fits2}) imply a rotational energy to X-ray energy
 conversion efficiency $L_x/\dot{E}\,$ of $\,1.1 \times 10^{-3}$
 within $0.5-10$ keV and of $3.4 \times 10^{-3}$ within the ROSAT
 band (cf.~Becker \& Tr\"umper 1997).

 In addition to the single component spectral models we tested two
 composite models consisting of either a blackbody plus power law
 component or of two blackbody components, respectively. The first
 model represents the scenario in which the X-ray emission of \PSRB\,
 originates from heated polar caps and from magnetospheric radiation
 processes.  The double blackbody model implies that its X-ray
 emission would be entirely of thermal origin, e.g.~with all the
 X-rays being emitted from heated polar caps with an anisotropic
 temperature distribution for which the model is approximated by two
 Planckian spectra with different emitting areas and temperatures.

 Clearly, the statistical motivation to include an additional thermal
 component to the already excellent power law fit is very small. All
 combinations of blackbody normalizations and temperatures that were
 fitted gave reduced $\chi^2$-values which are not better than the
 fits to a power law model of the pulsar spectrum. The F-test
 statistic for adding the extra blackbody spectral component to the
 power law model, thus, is very low. Based upon the errors of the
 fitted spectral components all fits of this composite model yield
 only upper limits for a maximum thermal component which the power law
 model ''accepts" before the fits become statistically
 unacceptable. The parameters obtained for the thermal component are
 thus intrinsically upper limits. The fitted parameters are shown in
 Tables~\ref{spectral_fits} and \ref{spectral_fits2}. The resulting
 blackbody temperature and the size of the projected emitting area are
 $kT \le 0.28$ keV and $R_{bb}\le 110$ m, assuming a pulsar distance
 of 361 pc.

\begin{figure}
\centerline{\psfig{figure=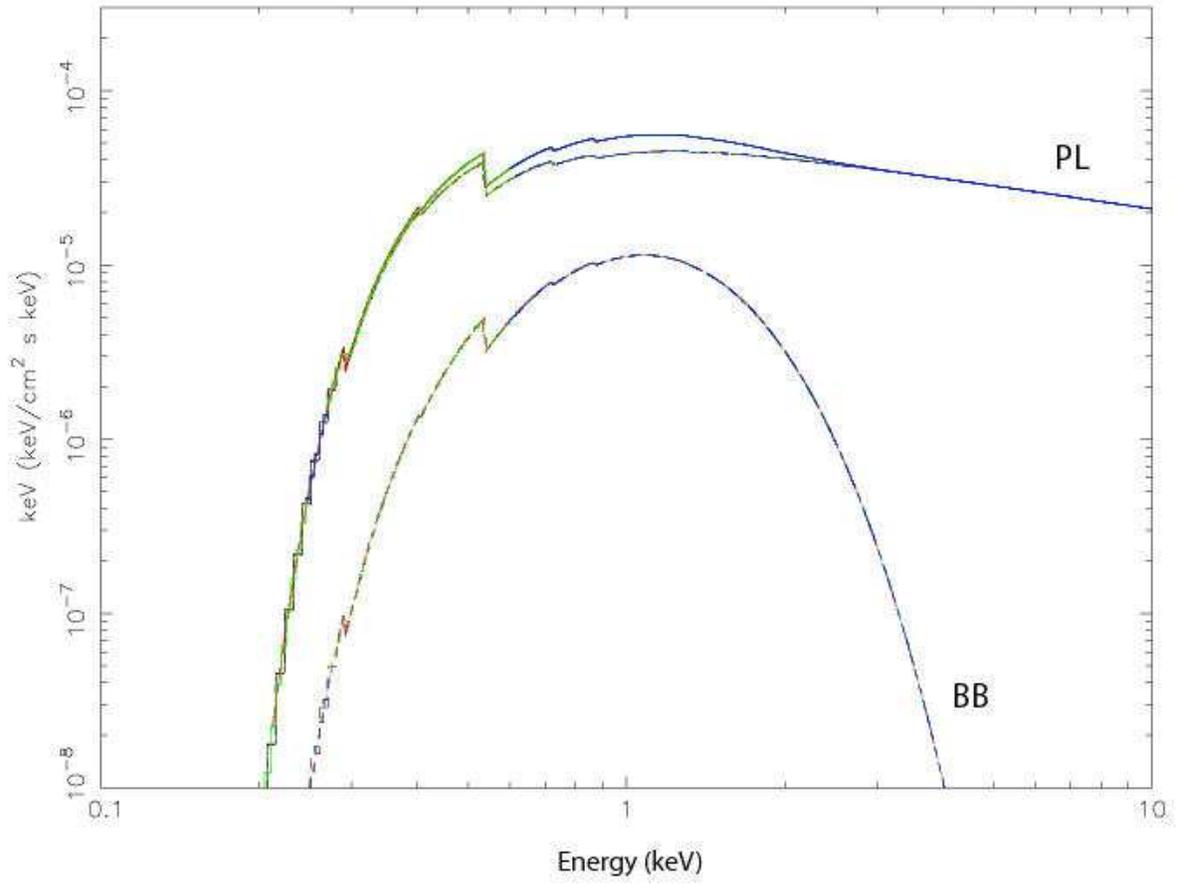,width=16cm,clip=}}
\caption[]{Blackbody plus power law spectral components and combined model as fitted to
the spectral data of \PSRB.}
\label{PSRB_bb_pl_model}
\end{figure}

 In computing the relative contributions of the thermal and
 non-thermal spectral components and stretching the errors to the
 limits for the composite blackbody plus power law model, we find that
 no more than $\sim 40\%$ of the detected X-ray flux could come from
 heated polar caps.  For the best fit parameters only $\sim 7\%$ could
 be emitted from these caps.  However, the radius of $\le 110$ m which
 we computed for the maximum projected emitting area is not too
 different from the expected size of a polar cap. For comparison,
 defining the size of the presumed polar cap as the foot points of the
 neutron star's dipolar magnetic field, the radius of the polar cap
 area is given by $\rho=\sqrt{2\pi R^3/c P}$ with $R$ being the
 neutron star radius, c the velocity of light and P the pulsar
 rotation period (see e.g.~Michel 1991). For \PSRB\, with a rotation
 period of 226 ms this yields a polar cap radius of $\rho\sim
 300\,\mbox{m}$. A plot of the thermal and non-thermal spectral
 components and the combined model is shown in Figure
 \ref{PSRB_bb_pl_model}.

 With  a spin-down  age of  $\sim 3\times  10^6$ years  \PSRB\, should
 still have some residual heat content from its birth event. Depending
 on the  equation of state the  surface temperature could  still be in
 the range  $\sim 1-3  \times 10^5$ K  (cf.~Becker \& Pavlov  2001 and
 references  therein) and  could  contribute  on a  low  level to  the
 detected soft  X-ray emission.  To estimate the  upper limit  for the
 surface  temperature   of  \PSRB\,   we  have  fixed   the  blackbody
 normalization so that the emitting area corresponds to the surface of
 a $R=10$ km neutron star  and calculated the confidence ranges of the
 blackbody temperature by leaving  all other parameters free.  We find
 a  $3\sigma$ surface  temperature upper  limit of  $T_s^\infty  < 4.5
 \times 10^5$ K which is  somewhat above the temperatures predicted by
 cooling models  (e.g.~Page \&  Applegate 1992; Tsuruta 1998; Yakovlev  
 et al.~1999) and, thus,  may constrain only  those thermal evolution  
 models which predict extreme reheating.

 The spectral model consisting of two thermal components describes the
 observed spectrum formally with comparable goodness of fit as the
 blackbody plus power law model.  In comparison with a single
 blackbody fit the F-test statistic thus supports an addition of a
 second thermal component to the single blackbody model. However,
 inspection of the fit residuals shows that the double blackbody model
 falls off rapidly beyond $\sim 5$ keV which causes the residuals
 beyond that energy to systematically lie above the zero line, albeit
 error bars are large. The fitted model parameters are a column
 absorption of $N_H=0.35_{-0.08}^{+0.11}\times
 10^{21}\,\mbox{cm}^{-2}$, temperatures $kT_1=0.59_{-0.05}^{+0.06}$,
 $kT_2=0.20_{-0.2}^{+0.2}$, and projected emitting areas
 $R_1=9.8_{-1.8}^{+2.1}\,\mbox{m}$ and $R_2=92_{-11}^{+14}\,\mbox{m}$,
 respectively.

 \subsubsection{PHASE RESOLVED SPECTRAL ANALYSIS\label{PSRB_phase_spec}}

 In order to investigate a possible variation of the pulsar emission 
 spectrum as a function of pulse phase $\phi$ we selected the events
 from phase intervals $\phi_1=]\,0.15- 0.65\,[$ and from $\phi_2
 =[\, 0.0 - 0.15; 0.65 - 1.0\,]$ (cf.~Fig.\ref{PSRB_x_radio_profiles}) 
 for a spectral analysis.
 The source counts are from a circle of $20''$ radius centered on the 
 target coordinates (the extraction region was smaller than the one 
 used for the phase-integrated case in order to improve the 
 signal-to-noise in the lower-statistics phase-resolved spectra). 
 The spectra were binned in order to have at least 30 counts per channel.
 Allowing both  the $N_{\rm H}$ and the photon index to vary we 
 obtained for the phase $\phi_1$ 
 $N_{\rm H}=(0.22\pm0.03)\times10^{22}$ cm$^{-2}$, $\Gamma=3.1\pm0.2$, 
 ($\chi^2_{\nu}=1.2$; 44 dof)  and for $\phi_2$ $N_{\rm H}=(0.17\pm0.03)
 \times10^{22}$ cm$^{-2}$, $\Gamma=2.7\pm0.2$ ($\chi^2_{\nu}=1.03$, 36  
 dof). Fixing the interstellar column to the phase-averaged value, the
 best fit values for the photon indices are $\Gamma=3.05\pm0.07$
 ($\chi^2_{\nu}=1.18$, 45 dof) and $\Gamma=3.00\pm0.09$
 ($\chi^2_{\nu}=1.05$, 37 dof) for the peak and off-pulse,
 respectively. Using composite (blackbody + power law) models 
 for both $\phi_1$ and $\phi_2$ did not yield better results.
 Within the statistical uncertainties there are therefore no spectral 
 changes observed as a function of pulse phase.

 \subsubsection{Diffuse emission from the X-ray trail of \PSRB \label{PSRB_spect_diff}}

 The large collecting area of XMM-Newton allows, for the first time, a
 spectral analysis of the emission from the pulsar's X-ray trail. The
 satellite roll-angle and the EPIC-PN reduced field of view of
 $4.4\times 4.4$ arcmin in SmallWindow mode caused only a small
 portion of the trail to be observed by the PN CCD \#4 in the short
 November 2003 observation. During the April 2004 observations the
 trail is outside the PNs field of view. Due to the increased
 instrument background towards the PN's readout node which is near to
 the location of the trail region in the CCD \#4 we did not include
 the PN data in the analysis of the diffuse emission. Both MOS cameras
 have covered a large portion of the trail in all observations
 (cf.~Figure \ref{PSRB_mos_fullfield}) so the analysis was restricted to
 the MOS1/2 data only.

 The photon statistic of the diffuse trail emission is sufficient
 for a detailed spectral modeling only near to the
 pulsar. We, therefore, extracted the X-ray spectrum from a circular
 region of 1 arcmin radius centered at RA(2000) $19^h\, 32^m\,
 07.5^s$, DEC $10^\circ\, 58'\, 32\arcsec$. The background spectrum 
 was extracted from a source-free region located at 
 RA(2000) $19^h\, 32^m\, 01.6^s$, DEC
 $10^\circ\, 59'\, 36\arcsec$.  About 800 counts ($\sim 60\%$
 background contribution) were available for the spectral analysis of
 the diffuse trail emission.  We binned the spectrum dynamically so as
 to have at least 30 counts per bin.

\begin{figure}
\centerline{\psfig{figure=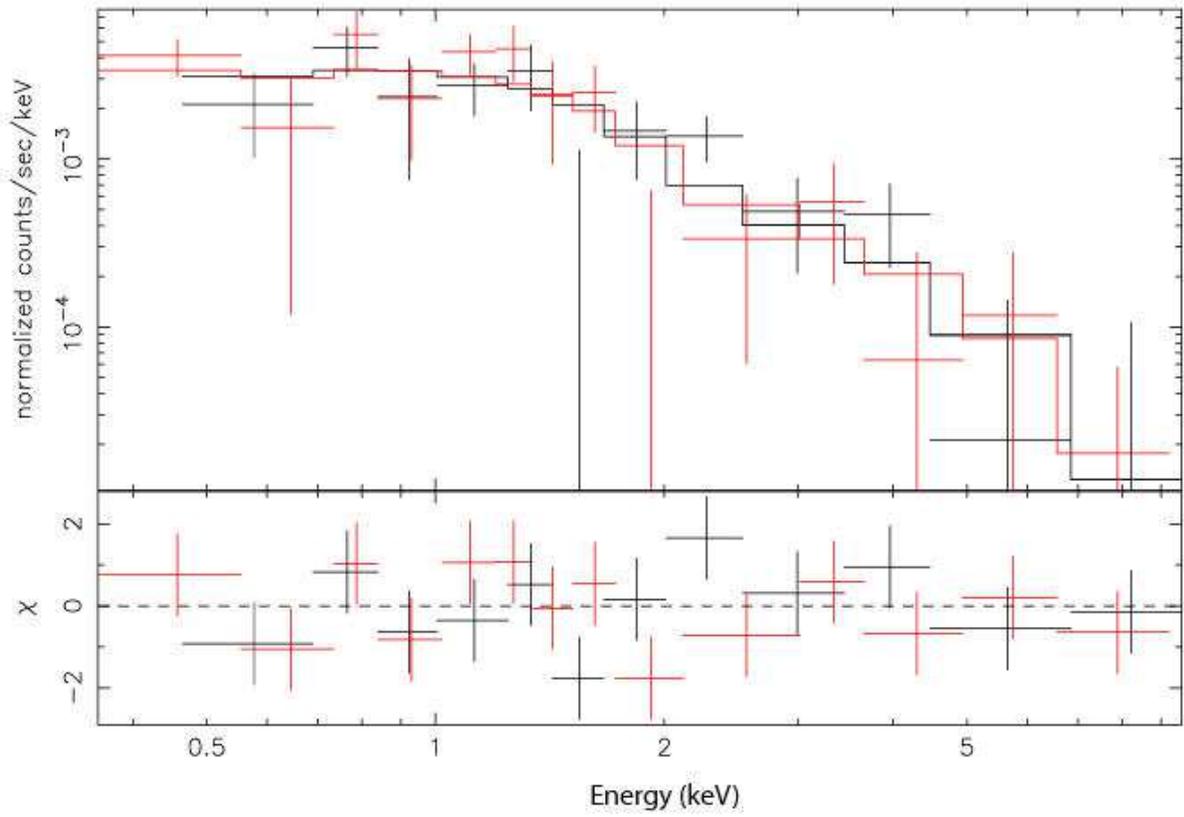,width=16cm,clip=}}
\caption[]{Energy spectrum of \PSRB's X-ray trail observed with the EPIC-MOS1/2
detectors and simultaneously fitted to an absorbed power law model ({\it upper panel})
and contribution to the \chisq\, fit statistic ({\it lower panel}).}
\label{PSRB_diffuse_pl_spectrum}
\end{figure}

\begin{figure}
\centerline{\psfig{figure=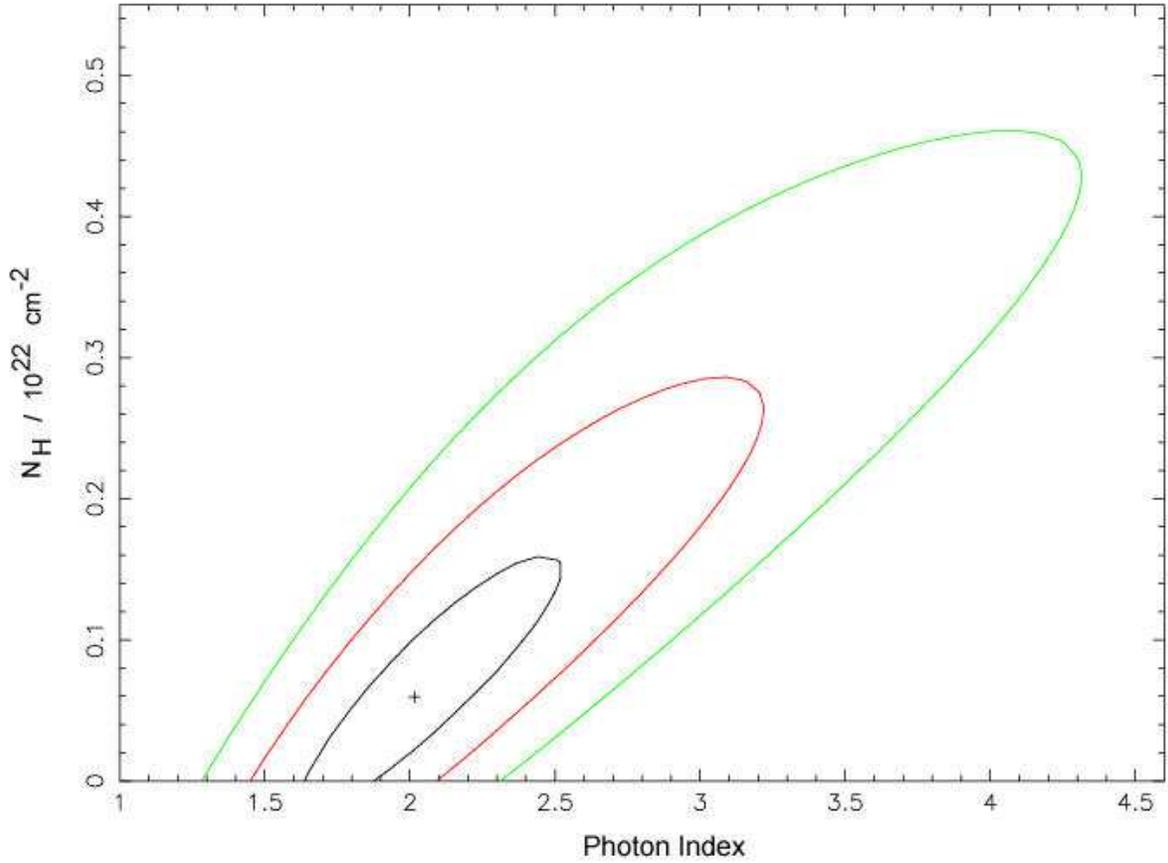,width=16cm,clip=}}
\caption[]{Contour plot showing the relative parameter dependence of the photon index vs.~column
absorption for the power law fit to the energy spectrum of \PSRB's X-ray trail. The three
contours represent the $1\sigma$, $2\sigma$ and $3\sigma$ confidence levels for one parameter
of interest. The `+' sign marks the best fit position, corresponding to $\chi^2_{min} =0.885$ for 23 dof.}
\label{PSRB_diffuse_pl_contour}
\end{figure}

 The pulsar's X-ray trail is likely formed by a ram-pressure confined
 pulsar wind.  Its X-ray emission should arise from synchrotron
 radiation of relativistic electrons with a spectral shape
 characterized by a power law. To test this hypothesis we fitted a
 power law to the extracted trail spectrum and found the model
 describes the observed spectrum well (\chisq =20.4 for 23 dof). The
 best-fit power law spectrum and residuals are shown in
 Figure~\ref{PSRB_diffuse_pl_spectrum}. Contour plots showing the
 relationship between the photon index and the column absorption for
 various confidence levels are shown in Figure
 \ref{PSRB_diffuse_pl_contour}.  Details of the spectral fits are
 again listed in Tables~\ref{spectral_fits} and \ref{spectral_fits2}.
 Accordingly, the 1 arcmin size portion of the diffuse nebula
 radiates $2.1 \times 10^{-4}\,\dot{E}$ into the $0.5 - 10$ keV X-ray 
 band and $2.3 \times 10^{-4}\,\dot{E}$ into the $0.1-2.4$ keV soft 
 X-ray band.

 For a second circular region which is located $\sim 2.5$ arcmin
 behind the pulsar along its proper motion direction we converted the
 background and vignetting corrected counting rate to an energy flux
 by assuming that the power law photon index is similar to the one we
 modeled from region one. This assumption may not be
 justified. For example, Willingal et al.~(2001) found in a
 detailed spectral analysis of the Crab nebula that its outer regions
 show the steepest spectrum.  This indicates enhanced synchrotron
 losses of the electrons during their passage from the pulsar to the
 outskirts of the nebula. It is conceivable that a
 similar behavior is valid for the electrons radiating in the trail of
 \PSRB. The energy fluxes and luminosities we obtain for the second
 region are upper limits in this respect.  For the $0.5-10$ keV and
 $0.1-2.4$ keV bands we find $f_x \sim 1.8 \times 10^{-14}\,\mbox{erg
 s}^{-1}\mbox{cm}^{-2}$, $f_x \sim 1 \times 10^{-14}\,\mbox{erg
 s}^{-1}\mbox{cm}^{-2}$ and $L_x \sim 2.8 \times 10^{29} \,\mbox{erg
 s}^{-1}$, $L_x \sim 1.6 \times 10^{29} \,\mbox{erg s}^{-1}$,
 respectively, which is $\sim 4-7 \times 10^{-5}\,\dot{E}$.

 Wang, Li \& Begelman (1993) investigated the trail emission by assuming 
 a synchrotron emission spectral shape for which they simulated a spectrum 
 and compared model predictions vs.~ROSAT observed counting rates. Our 
 findings based on a more detailed spectral modeling agree well with their 
 results.

 \subsection{MULTI-WAVELENGTH SPECTRUM\label{PSRB_multi_spec}}

\begin{figure}
\centerline{\psfig{figure=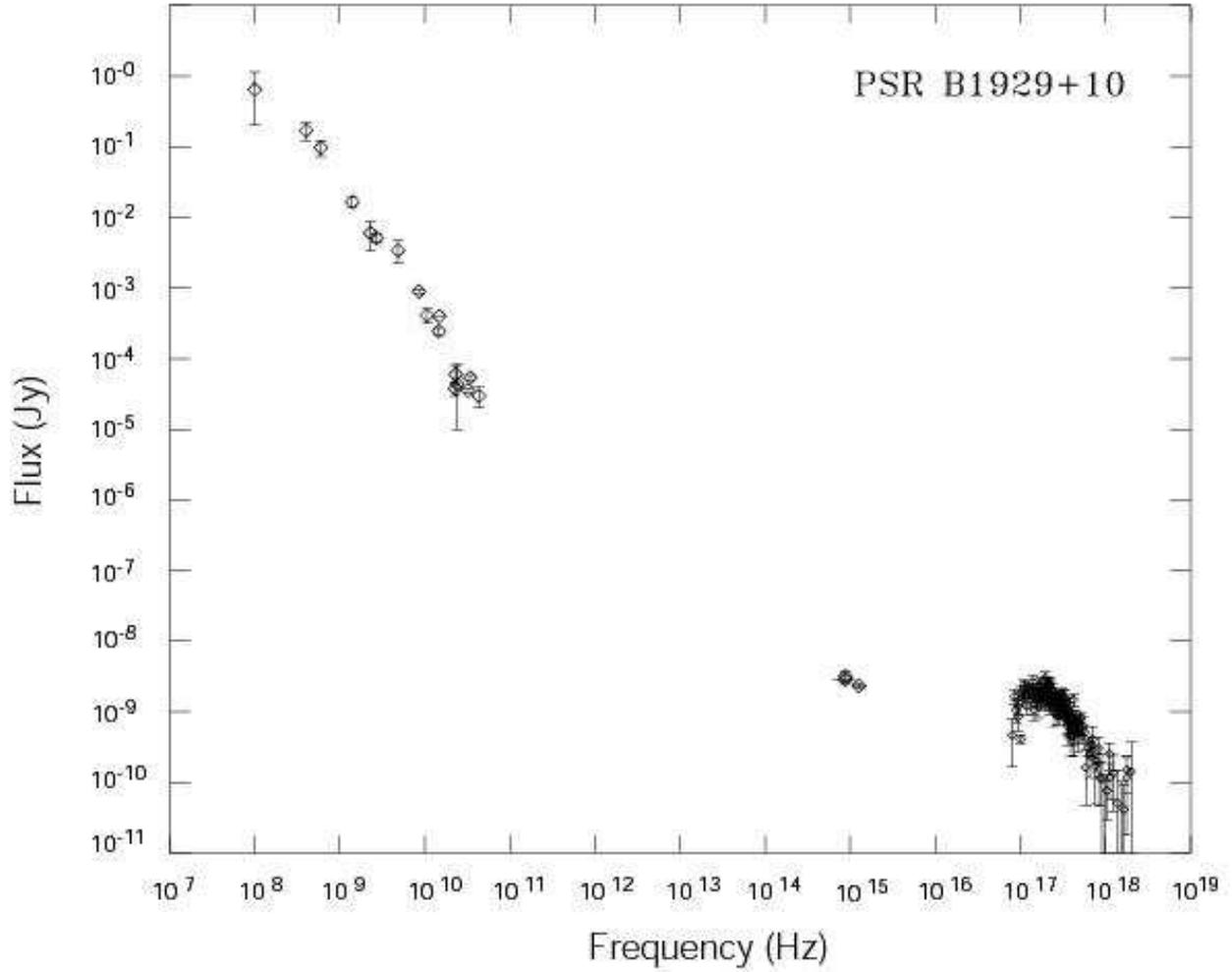,width=16.5cm,clip=}}
\caption[]{Combined radio, optical and X-ray spectral data of \PSRB. To describe the
 optical and X-ray spectrum in terms of magnetospheric emission a broken power law
 spectral model with a break point fitted at $E_{break}= 0.83^{+0.05}_{-0.03}\,\mbox{keV}$
 is required.} \label{PSRB_broadband_spectrum}
\end{figure}

 In order to construct a broadband spectrum combining all spectral
 information available from \PSRB\, we adopted the radio spectrum from
 Maron et al.~(2000) and plotted it in Figure
 \ref{PSRB_broadband_spectrum} together with the XMM-Newton observed
 pulsar spectrum and the optical fluxes obtained from the Hubble Space
 Telescope observations by Pavlov et al.~(1996) and Mignani et
 al.~(2002).

 \PSRB\, belongs to the small group of pulsars for which radio emission
 was detected up to 43~GHz (Kramer et al.~1997). The radio spectrum of
 \PSRB\, is a power law with a photon index of $\alpha=-2.6\pm 0.04$
 in the frequency range $0.4-24$~GHz (Maron et al.~2000). Its energy
 flux at 100~MHz is 950~mJy$\pm$ 600~mJy, but at 43~GHz it is still
 0.18~mJy~$\pm$~0.05~mJy (which is slightly more than extrapolated
 from the lower frequency data).  The flux density in the radio part
 of the spectrum, which should be due to coherent radiation, is
 therefore several orders of magnitude greater than the extrapolated
 optical or X-ray flux densities.

 Extrapolating the power law spectrum which describes the XMM-Newton
 data to the optical bands yields a photon flux which exceeds the flux
 measured in the near-UV bands by more than an order of magnitude.
 While the pulsar was clearly detected in the U-band (m$_{342W}=25.7$) 
 and with the F130LP (m$_{130LP}=26.9$) and STIS F25QTZ filters, only
 an upper limit is available in B (m$_{430W} \ge 26.2$) (Pavlov et 
 al.~1996; Mignani et al.~2002). This suggests that the broadband 
 spectrum, if entirely non-thermal, has to break somewhere before 
 or in the soft channels of the X-ray spectrum. To test this 
 hypothesis we have fitted a broken power law model to the XMM-Newton 
 and optical data. A broken power law model, with $E_{break}=
 0.83^{+0.05}_{-0.03}\,\mbox{keV}$, provides an excellent description
 (\chisq = 121.16 for 120 dof) of both spectral data sets. The photon 
 index for energies below and above the break $E_{break}$, is found 
 to be $\alpha_1=1.12^{+0.02}_{-0.03}$ and $\alpha_2 =2.48^{+0.08}_{-0.07}$,
 respectively (with a normalization of $8.2_{-0.7}^{+1.0}
 \times10^{-5}$ photons cm$^{-2}$ s$^{-1}$ keV$^{-1}$ at 1 keV). The
 model predicted column absorption value of $N_H=0.52_{-0.06}^{+0.12} \times
 10^{21}\,\mbox{cm}^{-2}$, somewhat smaller than the single power law
 model prediction, but is still compatible with our fit to the diffuse
 trail emission.

 \section{SUMMARY \& DISCUSSION\label{discussion}}

 We have investigated the emission properties of \PSRB\, and its X-ray
 trail in a multi-wavelength study using XMM-Newton, the ESO New
 Technology Telescope (NTT) in La Silla (Chile), the Hubble Space
 Telescope, the Effelsberg 100-m Radio Telescope and the Jodrell Bank
 Radio Observatory.

 In X-rays, \PSRB\, is the brightest pulsar among the old rotation-powered 
 pulsars. Therefore, for about 10 years it was the only member of its class 
 detected in X-rays and for which some details on its emission
 properties were known -- though this picture has been revised
 thoroughly in our multi-wavelength study of the old rotation-powered
 pulsars (cf.~Becker et al.~2004; 2005).

 Clearly, the study of old rotation-powered pulsars is in the domain
 of XMM-Newton which was designed and built to study faint sources in
 the X-ray sky. Its sensitivity allowed, for the first time, a more
 detailed study of this interesting group of pulsars.  As for PSR
 B0628-28 (Becker et al.~2005), PSR B0950+08, PSR B0823+26 and 
 PSR J2043+2740 (Becker et al~2004; Pavlov \& Zavlin 2004) recently
 observed by XMM-Newton, we found the temporal and spectral X-ray emission
 properties of \PSRB\, to be in excellent agreement with a non-thermal,
 magnetospheric emission scenario. Its X-ray
 spectrum is best described by a single power law model with a photon
 index of $2.72^{+0.12}_{-0.09}$. Using the best fit composite Planckian 
 power law model, the contribution from thermal emission of heated polar caps 
 is inferred to be at most $\sim 7\%$. However, a
 pure thermal emission spectrum consisting of two Planckian spectra is
 regarded as unlikely. Taking the optical spectral data into account a
 broken power law with $E_{break}= 0.83^{+0.05}_{-0.03}\,\,\mbox{keV}$
 and the photon-index $\alpha_1= 1.12^{+0.02}_{-0.03}$ and $\alpha_2
 =2.48^{+0.08}_{-0.07}\,$ is able to describe the emission in both
 spectral ranges entirely in terms of a non-thermal magnetospheric
 origin.

 The X-ray pulse profile observed in the $0.2-10$ keV band is found to
 be markedly different from the broad sinusoidal pulse profile seen in
 the low statistic ROSAT data. Fitting Gaussians to the X-ray light
 curve indicates the possible existence of three pulse components. A
 small narrow pulse, characterized by energies greater than 1 keV, is
 found to lead the radio main pulse by $\sim 20^{\circ}$. Two larger
 pulses, observed in all three energy bands, follow this small
 pulse. These three pulses are roughly separated by about the same
 phase cycle (cf.~Figure \ref{PSRB_three_comp_fit}). The fraction of 
 pulsed photons in the $0.2-10$ keV band is $32 \pm 4 \%$. For the 
 sub-bands $0.2-1.0$ keV and $1.0-2.1$ keV the pulsed fraction is 
 $24 \pm 5 \%$ and $44 \pm 6\%$, respectively, indicating a mild 
 energy dependence at a $\sim 2\sigma$ level.

 Various theoretical models have been developed to explain the observed 
 non-thermal high energy emission properties of younger pulsars like those
 in the Crab and Vela supernova remnants. They all appear to be seen not only 
 in X-rays but also in the gamma-ray band. Can the generic features of these
 models also explain the emission properties we observe from \PSRB?

 It is commonly believed that the non-thermal X-ray photons are
 emitted by relativistic charged particles in the pulsar
 magnetosphere. These relativistic particles could be accelerated in a
 polar cap region (e.g.~Harding 1981; Zhang \& Harding 2000) or in the
 outer-magnetosphere (e.g.~Cheng, Ho \& Ruderman 1986). In order to
 calculate the non-thermal spectrum and the energy dependent light
 curves, the amount of current flow and detailed three dimensional
 geometry of the accelerator are required. To fix these parameters,
 gamma-ray data is normally required. However, if the inclination and
 viewing angles of a specific pulsar are known, the qualitative
 features of the light curve can be predicted according to the three
 dimensional magnetospheric models (e.g. Yadigaroglu \& Romani 1995;
 Cheng, Ruderman \& Zhang 2000; Dyks, Harding \& Rudak 2004). 

\begin{figure}
\centerline{\psfig{figure=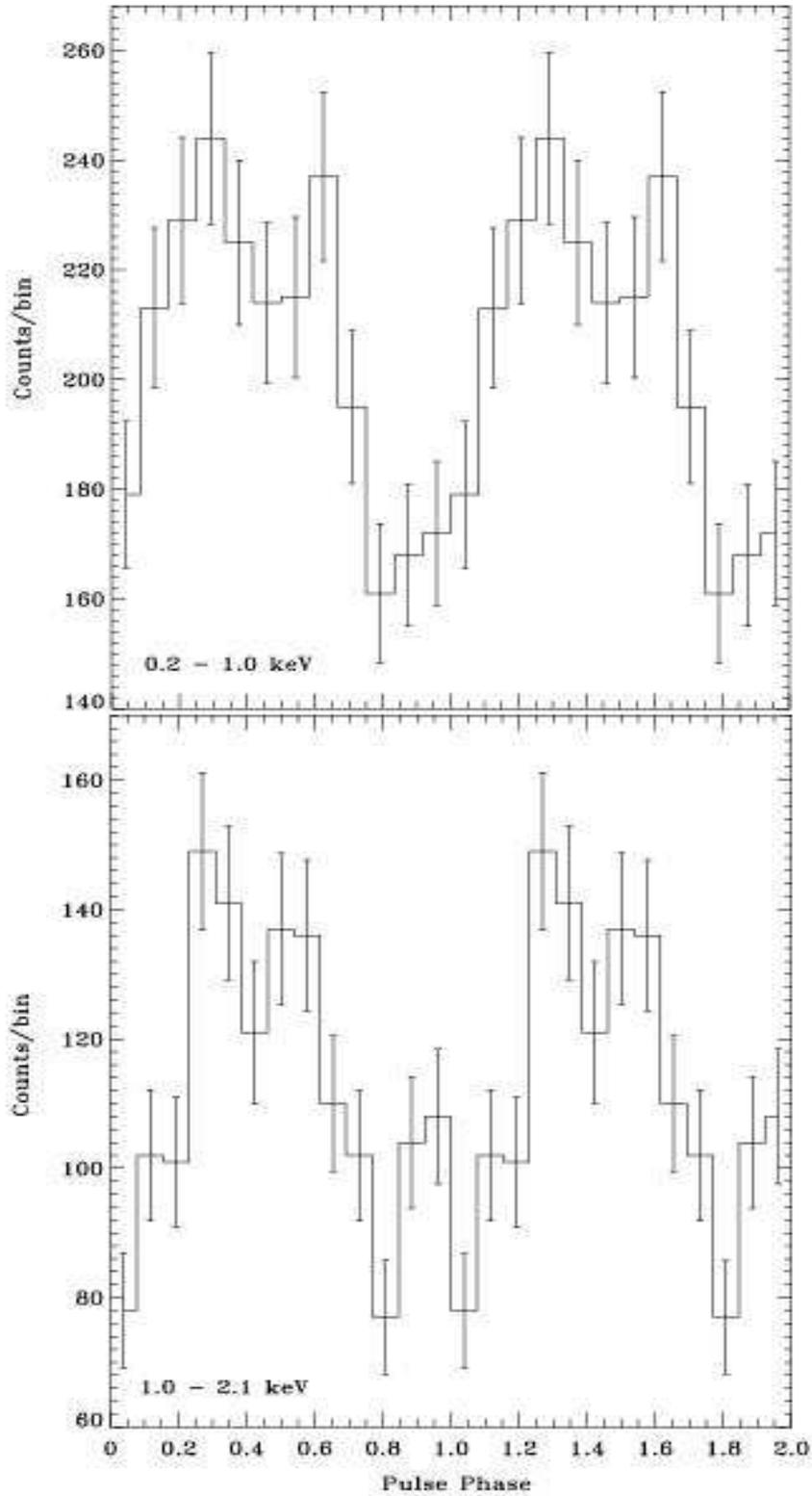,height=21cm,width=12cm,clip=}}
\caption[]{Integrated pulse profiles of \PSRB\, as observed in the energy bands $0.2-1.0$
keV and $1.0 - 2.1$ keV by the  EPIC-PN camera. A narrow pulse component appears beyond $\sim 1$ keV
near to phase 0 (1). Phase zero corresponds to MJD=53120.01448082337447 (TDB at SSB) and is the
location of the radio main pulse peak. Two phase cycles are shown for clarity. }
\label{PSRB_pulseprofiles}
\end{figure}

 Everett \& Weisberg (2001) have reported from radio polarization data 
 that the inclination and viewing angles of \PSRB\, are $36^{\circ}$ and
 $26^{\circ}$, respectively.  For such small inclination and viewing
 angles, at most two pulses with a phase separation of $\sim
 180^{\circ}$ can be produced if there is only outgoing current. In
 this case, one pulse arises from a region near one polar cap and
 another pulse arises from a region near the light cylinder, but
 associated with another magnetic pole. In Figure
 \ref{PSRB_pulseprofiles}, the first narrow peak only appears in the
 band $\sim 1.0-2.1$ keV and could be associated with a heated polar
 cap of kT $\sim$ 1keV. However, this is very unlikely by its
 narrowness as strong gravitational light bending would smear out the
 thermal pulse from the surface (e.g.~Page 1995). Cheng, Ruderman \&
 Zhang (2000) have shown that incoming current must exist from null
 charge surface to the stellar surface due to pair creation in the
 outer magnetosphere accelerator.

\begin{figure}
\centerline{\psfig{figure=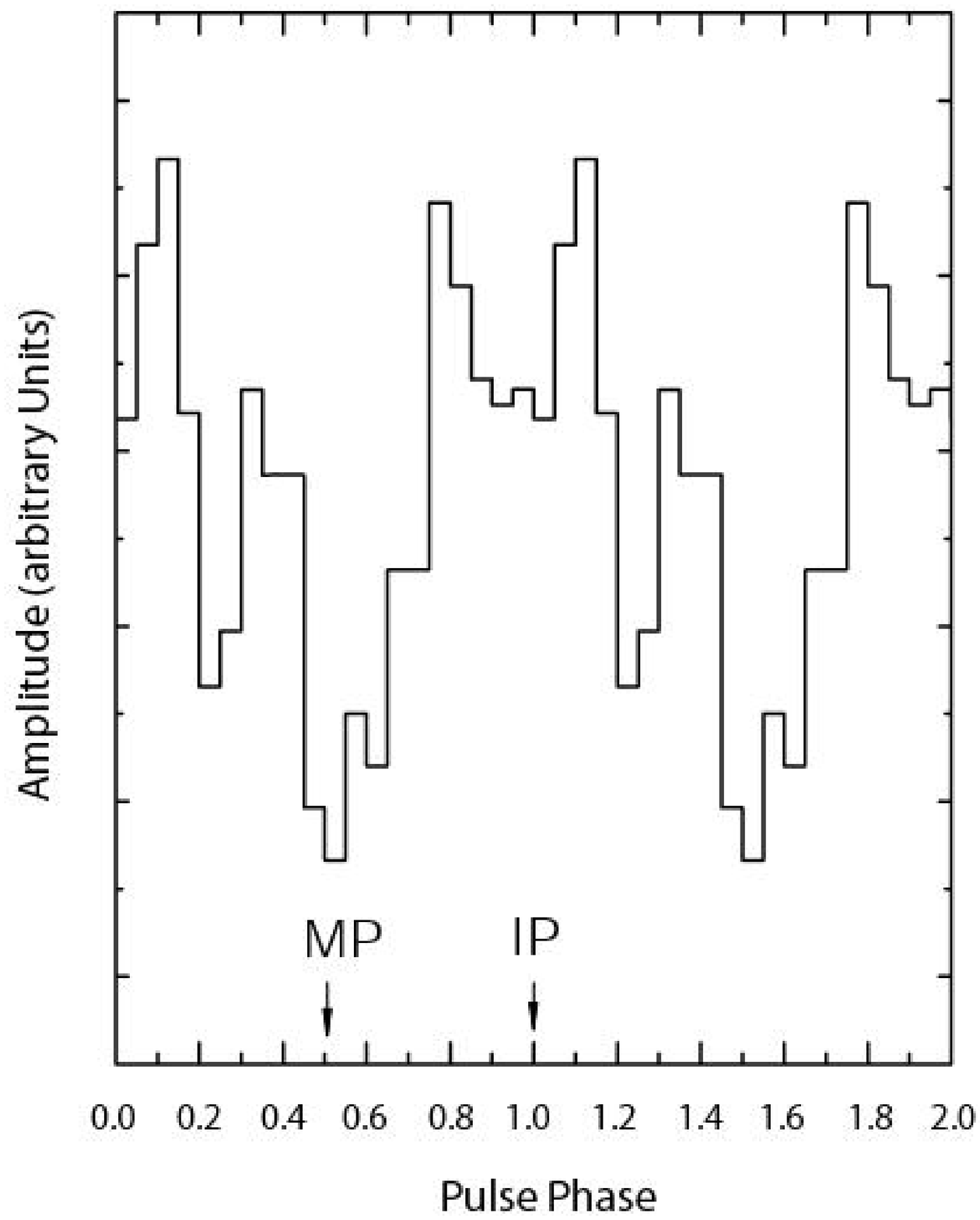,width=14cm,clip=}}
\caption[]{Simulated X-ray light curve of PSR B1929+10. Phase zero
refers to the plane containing both the magnetic and rotational
axes, and the related main pulse (MP) and interpulse (IP) of the
radio profile are indicated by arrows. Two phase cycles are
shown.} \label{sim_profile}
\end{figure}

 Here, we can simulate the light
 curves of \PSRB\ by using its observed inclination and viewing
 angles, and the three dimensional outer gap model (Cheng, Ruderman \&
 Zhang 2000). Since the radiation is expected to be emitted from open
 field lines, the coordinate values ($x_0, y_0, z_0$) of the last
 closed field lines at the stellar surface must be determined.  The
 coordinate values ($x_0^,, y_0^,, z_0^,$), where $x_0^,$=$a_1 x_0$,
 $y_0^, $=$a_1 y_0$ and $z_0^,$=$[1-(x_0^{,2} + y_0^{,2})]^{1/2}$,
 then represent an open field line surface for a given value of
 $a_1$. For simplicity, we choose $a_1 = 0.97$, which is very close to
 the first open field lines ($a_1=1.0$). In Figure \ref{sim_profile},
 we show that two large pulses and one small pulse can be
 simulated. The first large pulse is the result of the incoming current
 toward the south pole while another pulse is produced by the outgoing
 current from the north pole. The small pulse is produced by outgoing
 current near the light cylinder. We note that the phase separation of
 these three pulses roughly corresponds to the phase separation of the
 three pulse components observed in the X-ray light curve. Here, the
 (x,z) plane is chosen to lie at the zero-phase position in Figure
 \ref{sim_profile}. On the other hand, in Figure
 \ref{PSRB_x_radio_profiles}, the (x,z) plane should lie in the middle
 between the two radio pulses. Therefore, the simulated X-ray light
 curves roughly reproduce the phase relative to the radio pulses
 within a phase error of about $\pm 0.1$ which is well within the
 observed uncertainties.

\subsubsection*{THE TRAIL OF \PSRB}

 The existence of diffuse emission with a trail morphology lying
 in the direction opposite to the motion of the pulsar is confirmed
 in our XMM Newton observation and provides a unique opportunity to
 probe the pulsar and its environment.  The extended diffuse
 emission associated with this old pulsar (with spin down energy
 less than $10^{34}$ ergs s$^{-1}$) indicates that spin down
 power is not the sole criterion for its detection. In addition to
 distance, the detectability may also be dependent on the pulsar's
 transverse velocity (see Chatterjee \& Cordes 2002).  The existence
 of a possible radio counterpart in the 11 cm Effelsberg data is
 exciting and, if confirmed in subsequent observations, can provide
 important information on the trail properties.  No diffuse emission
 from the pulsar trail is detected in $H_\alpha$ perhaps suggesting that
 the neutral component of the interstellar medium is low (see Chatterjee
 \& Cordes 2002) in the environment surrounding \PSRB.

 Since the trails X-ray emission near to the pulsar has a hard spectrum 
 characterized by a power law photon index of $\sim 2$, the emission is 
 non-thermal and is likely to be produced from the synchrotron process of
 highly relativistic electrons in the shocked region between the pulsar 
 wind and surrounding interstellar medium. A physical description of 
 \PSRB's X-ray trail based on the ROSAT findings has been discussed 
 by Wang, Li, \& Begelman (1993) in terms of an outflow collimated 
 within the pulsar's cavity created by its motion.

 Alternatively, the properties of the distorted wind nebula can be
 inferred under the assumption that the electron lifetime due to
 synchrotron losses, $\tau_{syn}$, is comparable to the timescale for
 the passage of the pulsar over the length of its X-ray trail. Such a
 hypothesis has also been considered, for consistency, by Caraveo et
 al.~(2003) in their model analysis of a similar X-ray trail observed
 in the Geminga pulsar.  For PSR 1929+10, the angular extent of the
 trail can not be definitively determined with the present data, but 
 it is likely greater than 4 arcmin. 
 For an assumed distance of 361 pc, the linear scale, $d$, corresponding 
 to this angular scale is greater than about 0.4 pc.  An estimate of the
 flow time, $t_{flow}$, can be obtained, for the proper motion measured 
 by Chatterjee et al.~(2004) and resulting velocity of $v_p$ of 177 km 
 s$^{-1}$, leading to  $t_{flow} \gtrsim 2200$ yrs. This lower limit is 
 about a factor of 2.3 times longer than the comparable timescale found
 for the Geminga pulsar (see Caraveo et al.~2003). To determine the
 consistency of our interpretation, the magnetic field in the shocked
 region can be estimated by equating $\tau_{syn}$ to $t_{flow}$.  Here
 $\tau_{syn}= 6\pi m_ec/\gamma \sigma_T B^2 \sim 10^5 B_{\mu G}^{-3/2}
 (h\nu_X/ {\rm keV})^{-1/2}$ yr where $\gamma$ is the Lorentz factor
 of the wind, taken to be equal to $10^6$, $\sigma_T$ is the Thompson
 cross section, and $B_{\mu G}$ is the magnetic field in the emission
 region in micro gauss.  The inferred magnetic field strength in the
 emitting region is $\lesssim 12 \mu$G. Given the magnetic field strength
 estimates in the interstellar medium ($\sim 2 - 6 \mu$G; see Beck et
 al. 2003), and the expected compression of the field in the
 termination shock by about a factor of 3 (Kennel \& Coroniti 1984),
 our estimates for the magnetic field in the emitting region of the
 pulsar wind nebula are in approximate accord.

 The X-ray luminosity and the spectrum of the emitted radiation can be
 estimated using a simple, one-zone model for the emission nebula
 powered by the pulsar wind as developed by Chevalier (2000). To
 determine the characteristic properties of the emission a comparison
 of the cooling frequency, $\nu_c$, for which the electrons can
 radiate their energy in the pulsar trail, with the observing
 frequency, $\nu_X$ is necessary. The cooling frequency can be
 expressed as $\nu_c= {e\over 2\pi m_ecB^3} ({6\pi m_ec\over \sigma_T
 \tau_{syn}})^2$. Substitution of the inferred magnetic strength and
 the electron cooling timescale due to synchrotron radiation yields
 $\nu_c = 1.8 \times 10^{17}$ Hz.  Since $\nu_X > \nu_c$, the
 electrons are able to radiate their energy in the pulsar trail and
 the photon index, $\Gamma$, is related to the power law index, $p$,
 of the electron energy distribution, $N(\gamma)\propto \gamma^{-p}$,
 in the form $\Gamma=(p+2)/2$.  Based on the theoretical work on
 highly relativistic shocks (Bednarz \& Ostrowski 1998, Lemoine \&
 Pelletier 2003), we adopt $p=2.2$, yielding $\Gamma = 2.1$. To be
 consistent with our interpretation of emission taking place in the
 fast cooling region, the observed value of $\Gamma$ should exceed 2,
 which is consistent with the observed value of $\Gamma
 =2_{-0.4}^{+0.4}$.

 The luminosity of the radiating electrons in the nebula can be
 calculated from the luminosity per unit frequency given by $$L_\nu =
 {1\over 2}({p-2\over p-1})^{p-1} ({6e^2\over 4\pi^2 m_e
 c^3})^{(p-2)/4} \epsilon_e^{p-1}\epsilon_B^{(p-2)/4} \gamma_w^{p-2}$$
 \beq R_s^ {-(p-2)/2}{\dot E}^{(p+2)/4}\nu^{-p/2}. \enq where $R_s$ is
 the distance of the shock from the pulsar expressed as $R_s = ({\dot
 E}/2\pi \rho v_p^2 c)^{1/2}\sim 10^{16}{\dot
 E}_{33}^{1/2}n^{-1/2}v_{p,100}^{-1} {\rm cm}$. Here $v_{p,100}$ is
 the velocity of the pulsar in units of 100 km s$^{-1}$, $\dot E_{33}$
 is the spin down power of the pulsar in units of $10^{33}$ ergs
 s$^{-1}$, and $n$ is the number density of the interstellar medium in
 units of 1 cm$^{-3}$.  The spin down power of $3.89 \times 10^{33}$
 ergs s$^{-1}$ and a density of 1 cm$^{-3}$ yields a shock radius of
 $R_s \sim 6 \times 10^{15}$ cm.  Assuming energy equipartition
 between the electron and proton fractional energy densities so that
 $\epsilon_e \sim 0.5$, and a fractional energy density of the
 magnetic field $\epsilon_B \sim 0.01$ (see Cheng, Taam, \& Wang
 2004), the corresponding luminosity given as $\nu L_\nu$ is $ \sim
 1.6 \times 10^{30}$ ergs s$^{-1}$ or $\sim 4 \times 10^{-4} \dot
 E$. In view of the observational uncertainties, this is consistent
 with the observed values of $8.3_{-3.4}^{+8.7} \times 10^{29}$ ergs
 s$^{-1}$ ($0.5-10$ keV) and $9.1_{-3.6}^{+8.6} \times 10^{29}$ ergs
 s$^{-1}$ ($0.1 - 2.4$ keV).

 Future theoretical investigations should be carried out to confronting 
 the observed X-ray lightcurves and spectra having even better photon 
 statistics than the ones which we have obtained in this first XMM-Newton 
 observations of \PSRB. The results we obtained from the X-ray trail along 
 with the possible discovery of its radio counterpart are exciting. Follow 
 up radio observations of the trail region at different wavelengths are 
 currently scheduled for fall 2006 in order to further constrain its existence 
 and, if detected, provide polarization information and the spectral index 
 in the radio regime from it.

\acknowledgments

We acknowledg he use of data obtained from XMM-Newton observations which is an 
ESA science mission with instruments and contributions directly funded by ESA Member 
States and NASA. WB would like to thank Hermann Brunner for his help in the X-ray image 
reconstruction. We are grateful to Olaf Maron for the use of his database of pulsar 
radio fluxes, to E.~F\"urst for help with the 11cm Effelsberg radio survey and to
Armando Manzali for his contribution to the the phase resolved spectral analysis 
which is part of the thesis work. KSC and JJJ are partially supported by a RGC 
grant of Hong Kong Government. AS was supported by KBN grant PBZ-KBN-054/P03/2001. 
ADL acknowledges a fellowship by the Italian Space Agency, ASI. PAC and ADL 
acknowledge financial contribution from contract ASI-INAF I/023/05/0. Optical 
observations were made with ESO Telescopes at the La Silla Observatories 
under program 077.D-0794(A). This work was supported in part by the Theoretical 
Institute for Advanced Research in Astrophysics (TIARA) operated under Academia 
Sinica and the National Science Council Excellence Projects program in Taiwan 
administered through grant number NSC 94-2752-M-007-001.

\clearpage

\begin{deluxetable}{lc}
\tablewidth{0pc}
\tablecaption{Ephemerides of \PSRB \label{t:radio}}
\tablehead{}
\startdata

Right Ascension (J2000)                                          \dotfill      & $19^h 32^m 13^s\!.983 \pm 00^s\!.002$  \\
Declination (J2000)                                              \dotfill      & $+10^d 59^m 32^s\!.41 \pm 00^s\!.07$    \\
First date for valid parameters (MJD)                            \dotfill      & 52929                            \\
Last date for valid parameters (MJD)                             \dotfill      & 53159                            \\
Infinite-frequency geocentric pulse arrival time$^a$ (MJD, UTC)  \dotfill      & 53120.000002393                  \\
Pulsar rotation period ($s$)                                     \dotfill      & 0.2265182954                     \\
Pulsar rotation frequency ($s^{-1}$)                             \dotfill      & 4.4146544466756                  \\
First derivative of pulsar frequency ($s^{-2}$)                  \dotfill      & $-2.26998 \times 10^{-14}$       \\
Second derivative of pulsar frequency ($s^{-3}$)                 \dotfill      & $2.72 \times 10^{-26}$           \\
Spin-down age (yr/$10^6$)                                        \dotfill      & 3.09                             \\
Spin-down energy ($\mbox{erg s}^{-1}/10^{33}$)                        \dotfill      & 3.89                             \\
Inferred Magnetic Field ($G/10^{12}$)                            \dotfill      & 0.5129                           \\
Dispersion Measure ($\mbox{pc cm}^{-3}$)                            \dotfill      & 3.178                            \\
Distance$^b$ (pc)                                                \dotfill      & $361_{-8}^{+10}$                 \\
\enddata
\tablecomments{\newline
$^a$ The integer part of this time is the barycentric (TDB) epoch of RA, DEC, f, $\dot{f}, \ddot{f}$.\newline
$^b$ Distance based on radio parallax according to Chatterjee et al (2004).}
\end{deluxetable}

\clearpage

\begin{deluxetable}{lcr}
\tablewidth{0pc}
\tablecaption{Details of the XMM-Newton observations of \PSRB.\label{t:xray_obs}}
\tablehead{}
\startdata
 Detector  &   Duration    &   eff.~Exposure \\
    {}     &     sec        &       {sec\quad\quad}   \\\hline\\[-1ex]

\multicolumn{3}{c}{SEQ: 0718\_0113051{\bf 301} / 2003-11-10}\\\\[-2ex]
\quad EMOS1 &       10677     &  7345.1\quad\quad   \\
\quad EMOS2 &       10668     &  7893.3\quad\quad   \\
\quad EPN   &       10470     &  6994.0\quad\quad  \\\hline\\[-1ex]
\multicolumn{3}{c}{SEQ: 0718\_0113051{\bf 401} / 2004-04-27}\\\\[-2ex]
\quad EMOS1 &       21665     & 13499.2\quad\quad \\
\quad EMOS2 &       21676     & 14351.9\quad\quad \\
\quad EPN   &       21470     & 10556.0\quad\quad \\ \hline\\[-1ex]
\multicolumn{3}{c}{SEQ: 0718\_0113051{\bf 501} / 2004-04-29}\\\\[-2ex]
\quad EMOS1 &       22673     & 10271.6\quad\quad  \\
\quad EMOS2 &       22678     &  6353.6\quad\quad  \\
\quad EPN   &       22471     &  6268.3\quad\quad  \\ \hline
\enddata

\tablecomments{\small MOS1/2 observations were performed in FullWindow mode while
 the PN camera was setup to operate in SmallWindow mode. The
 medium filter was used in all observations.}

\end{deluxetable}

\clearpage

\begin{deluxetable}{ccccccc}
\tablewidth{0pc}
\tablecaption{Models and parameters as fitted to the energy spectrum of \PSRB\, and its X-ray trail.
\label{spectral_fits}}
\tablehead{}
\startdata
 model$^a$ & $\chi_\nu^2$ & $\nu$   &      $N_H/10^{21}$    &   $kT^b\, / \,\alpha$    &     Normalization at 1 keV           &  Radius$^c$  \\
  {}       &      {}      &   {}    &    $\mbox{cm}^{-2}$   &       {}                 &     Photons/keV/cm${^2}$/s           &      m       \\\hline\\[-1ex]

  PL       &     0.989    &  121    & $1.6_{-0.18}^{+0.2}$  &  $2.72_{-0.09}^{+0.12}$  &   $8.1_{-0.5}^{+0.6}\times 10^{-5}$  &      {}       \\\\[-0.1ex]

  BB+PL    &     0.997    &  119    & $1.1_{-0.3}^{+0.6}$   & $ \le 0.28 \,
                                                              /\, 2.45_{-0.21}^{+0.33}$&  $5.8_{-1.4}^{+1.7}\times 10^{-5}$    &   $\le 106$  \\\\[-0.1ex]

  BB+PL    &     0.996    &  120   & $1.6_{-0.2}^{+0.3}$   & $\le 0.04 \,
                                                             /\, 2.73_{-0.1}^{+0.13}$     & $8.2_{-0.6}^{+0.8}\times 10^{-5}$  &   $10\,000$  \\\\[-0.1ex]

BKNPL$^d$  &      1.010    &   120   & $0.52_{-0.06}^{+0.12}$& \parbox{3cm}
                                                              {$\alpha_1=1.12_{-0.03}^{+0.02}$\\[1ex]
                                                              $\alpha_2=2.48_{-0.07}^{+0.08}$}
                                                                                        & $8.2_{-0.7}^{+1.0}\times 10^{-5}$    &        {}    \\\\[-0.1ex]

  BB+BB   &     1.069     &  119    &$0.35_{-0.08}^{+0.11}$ &\parbox{3cm}{$\mbox{kT}_1=
                                                            0.59_{-0.05}^{+0.06}$\\[1ex]
                                                            $\mbox{kT}_2=
                                                                0.20_{-0.2}^{+0.2}$ }  &                {}                     & \parbox{0.9cm}
                                                                                                                                  { $9.8_{-1.8}^{+2.1}$ \\[1ex]
                                                                                                                                   $\quad92_{-11}^{+14}$} \\[4ex]\hline\\

\multicolumn{7}{c}{Diffuse emission from the pulsar's X-ray trail}\\\\[0ex]

 PL       &     0.8853    &  23    & $0.6_{-0.6}^{+0.7}$  &  $2.0_{-0.4}^{+0.4}$  &   $1.2_{-0.3}^{+0.5}\times 10^{-5}$  &      {}       \\[2ex]\hline
\enddata
\tablecomments{\newline
$^a$ BB = blackbody,  PL = power law,  BKNPL = broken power law\\
$^b$ The entry in this column depends on the spectral model --- it is the temperature $kT$ in keV or the photon index $\alpha$.\\
$^c$ For thermal models for which we computed or fixed the radius of the emitting area we assumed a pulsar distance of 361 pc.\\
$^d$ The broken power law model has been fitted to the combined X-ray and optical spectral data. The break energy is
fitted to be $E_{break}=0.83_{-0.03}^{+0.05}$ keV }
\end{deluxetable}

\clearpage

\begin{deluxetable}{cccc}
\tablewidth{0pc}
\tablecaption{Unabsorbed energy flux and X-ray luminosity derived from various modeled
fitted to the energy spectrum of \PSRB\, and its X-ray trail.
See Table~\ref{spectral_fits} for details.
\label{spectral_fits2}}
\tablehead{}
\startdata
 model     &  Energy band & $f_x$ & $L_x$ \\
  {}       & keV  & $10^{-13}$ erg s$^{-1}$ s$^{-2}$ & $10^{30}$ erg s$^{-1}$    \\\hline\\[-1ex]

  PL       &    $0.5-10$  &  $2.64^{+0.12}_{-0.16}$ & $4.11^{+0.18}_{-0.16}$      \\
           &    $0.1-2.4$ &  $8.59^{+2.17}_{-1.36}$ & $13.4^{+3.4}_{-2.1}$      \\\\[-0.1ex]

  BB+PL    &    $0.5-10$   &  $2.37^{+4.88}_{-0.56}$ & $3.69^{+7.6}_{-0.88}$      \\
           &    $0.1-2.4$  &  $4.7^{+9}_{-2}$ & $7.38^{+14.3}_{-3.1}$      \\\\[-0.1ex]

\multicolumn{4}{c}{Diffuse emission from the pulsar's X-ray trail}\\\\[0ex]

 PL       &     $0.5-10$   &  $0.53^{+0.56}_{-0.22}$ & $0.83^{+0.87}_{-0.34}$      \\
          &    $0.1-2.4$  &  $0.58^{+0.55}_{-0.23}$ & $0.91^{+0.86}_{-0.36}$      \\\\[-1ex]\hline

\enddata
\end{deluxetable}

\clearpage

\clearpage

\clearpage

\clearpage

\clearpage

\clearpage

\clearpage

\clearpage

\clearpage

\clearpage

\clearpage

\clearpage

\clearpage

\clearpage

\clearpage

\clearpage

\begin{thebibliography}

\bibitem[]{}Beck, R., Shukurov, A., Sokoloff, D., \& Wielebinksi, R. 2003. \aap, 411, 99

\bibitem[]{} Becker, W., Jessner, A., Kramer, M., Testa, V., Howaldt C., 2005, ApJ, 633, 367

\bibitem[]{} Becker, W., Weisskopf, M.C., Tennant, A.F., Jessner, A., Dyks,  J., Harding, A.K., Zhang, S.N., 2004, ApJ, 615, 908

\bibitem[]{} Becker, W., Aschenbach, B., 2002, in {Proceedings of the
   WE-Heraeus Seminar on Neutron Stars, Pulsars and Supernova remnants},
   Eds.~W.Becker, H.Lesch \& J.Tr\"umper, MPE-Report 278, 64, (available from astro-ph/0208466)

\bibitem[]{} Becker, W., Pavlov, G.G., 2001, in {\em The Century of Space
   Science}, Eds.~J.Bleeker, J.Geiss \& M.Huber, Kluwer Academic Publishers, p721 (available from astro-ph/0208356).

\bibitem[]{} Becker, W. \& Tr\"umper, J. 1999, A\&A, 341, 803

\bibitem[]{} Becker, W., Tr\"umper, J., 1997, A\&A, 326, 682

\bibitem{} {}Bednarz, J. \& Ostrowski, M. 1998, Phys. Rev. Lett., 80, 3911

\bibitem[]{} Briel, U.G., Burkert, W., Pfeffermann, E., 1989, in {\em EUV, X-ray, and Gamma-ray
   Instrumentation for Astronomy and Atomic Physics}, eds C.J.~Hailey \& O.H.W.~Siegmund, SPIE Vol. 1159, 263

\bibitem[]{} Buccheri R., De Jager O.C., in Timing Neutron Stars, Eds.~H.\"Ogelman,
   E.P.J.~van den Heuvel, p95, Kluwer Academic Publishers, 1989

\bibitem{} {}Caraveo, P. A., Bignami, G. F., DeLuca, A., Mereghetti, S., Pellizzoni, A., Mignani, R., Tur, A., \& Becker, W. 2003, Sci., 301, 1345

\bibitem[]{} Chatterjee, S., Cordes, J.M., Vlemmings, W.H.T., Arzoumanian, Z., Goss, W.M., Lazio, T.J.W., ApJ, 2004, 604, 339

\bibitem[]{} Chatterjee, S., Cordes, J.~M., 2002, ApJ, 575, 407

\bibitem[]{} Cheng, K. S., Ho, C., Ruderman, M. 1986, ApJ, 300, 500

\bibitem[]{} Cheng, K. S., Ruderman, M. A., Zhang, L. 2000, ApJ, 537, 964

\bibitem[]{} Cheng, K. S., Taam, R. E., Wang, W., 2004, ApJ, 617, 480

\bibitem[]{} Chevalier, R. A., 2000, ApJ, 539, L45

\bibitem[]{} Chiang, J., \& Romani, R. W., 1992, ApJ, 400, 629

\bibitem[]{} Condon, J.J., Cotton, W.D., Greisen, E.W., Yin, Q.F., Perley, R.A., Taylor, G.B., Broderick, J. J. 1998, AJ, 115, 1693

\bibitem[]{} Cordes, J.M., Lazio, T.J.W., 2002, astro-ph/0207156

\bibitem[]{} De Jager O.C., 1987, thesis, Potchefstromm University for Christian Higher Education, South Africa

\bibitem[]{} De Luca, A., Caraveo, P.A., Mereghetti, S., Negroni, M., Bignami, G.F., 2005, ApJ, 623, 1051

\bibitem[]{} Downs, G.S., Reichley, P.E., 1983, ApJS, 53,169

\bibitem[]{} Dyks, J., Harding, A. K., \& Rudak, B. 2004, ApJ, 606, 1125

\bibitem[]{} Dyks, J., \& Rudak, B. 2004, AdSpR, 33, 581

\bibitem[]{} Everett, J.E., Weisberg, J.M., 2001, ApJ, 553, 341

\bibitem[]{} Harding, A.K., Muslimov, A.G., 2003, in the proceedings of
  {\em Pulsars, AXPs and SGRs Observed by BeppoSAX and Other Observatories}, astro-ph/0304121

\bibitem[]{}  Harding, A.K., Muslimov, A.G.,2002, ApJ, 568, 862

\bibitem[]{}  Harding, A.K., Muslimov, A.G.,2001, ApJ, 556, 1001

\bibitem[]{} Harding, A. K. 1981, ApJ, 245, 267

\bibitem[]{} Helfand, D.J., 1983, in {\em Supernova Remnants and Their X-ray Emission},
  Eds.~J.Danziger and P.Gorenstein, Proceedings of the IAU Symposium No.~101, p471

\bibitem[]{} Hobbs, G., Lyne, A. G., Kramer, M., Martin, C. E., Jordan, C. A., 2004, MNRAS, 353, 1311

\bibitem{} {}Kennel, C.F.,  Coroniti, F.V. 1984, \apj, 283, 694

\bibitem[]{} Kirsch, M.G.F., Becker, W., Benlloch-Garcia, S., et al., in {\em X-Ray and Gamma-Ray
  Instrumentation for Astronomy XIII}, eds Flanagan, K.A. \& Siegmund, O.H.W., SPIE, Volume 5165, 85, 2004

\bibitem[]{} Kramer, M., Jessner, A., Doroshenko, O., Wielebinski, R., 1997, A\&A, 488, 364

\bibitem[]{} Kramer, M.,  Wielebinski, R., Jessner, A., Gil, J.A., Seiradakis, J.H., 1994,
  Astr. \& Astrophys. Suppl., 107, 515

\bibitem[]{} Krautter, J., Zickgraf, F.-J., Appenzeller, I., Thiering, I., Voges, W.,
   Chavarria, C., Kneer, R., Mujica, R., Pakull, M.W., Serrano, A., Ziegler, B., 1999, A\&A, 350, 743

\bibitem[]{} Kouwenhoven, M.L.A., van den Berg, M.C., 2001, A \& A, 367, 931

\bibitem[]{}Lemoine, M., Pelletier, G. 2003, \apj, 589, L73

\bibitem[]{}  Manning, R., Willmore, P., 1994, MNRAS, 266, 635

\bibitem[]{}  Maron, O., Kijak, J., Kramer, M., Wielebinski, R., Astron.~Astrophys.~Suppl.~Ser.~147, 195, 2000


\bibitem[]{} Michel, F.C., 1991, {\em Theory of Neutron Star Magnetospheres}, University of Chicago Press, Chicago, IL

\bibitem[]{} Mignani, R.P., De Luca, A., Caraveo, P.A., Becker, W., ApJ, 2002, ApJ, 580L, 147

\bibitem[]{} Page, D. 1995, ApJ, 442, 273

\bibitem[]{} Page, D., Applegate, J.L. 1992, ApJ, 394, L17

\bibitem[]{} Pavlov, G.G., Stringfellow, G.S., Cordova, F.A., 1996, ApJ, 467,370

\bibitem[]{}  Reich, W., Fuerst, E., Reich, P., Reif, K., 1990, A\&A, Suppl.~85, 633

\bibitem[]{} Saito, Y., 1998, PhD Thesis, ISAS Research Note 643

\bibitem[]{} Sembay, S., Abbey, A., Altieri, B., Ambrosi, R., Baskill, D., 
Ferrando, P., Mukerjee, K., Read, A.M., Turner, M.J.L., 2004, Proceedings of the SPIE, 5488, 264

\bibitem[]{} Seward, F.D., Wang, Z.R., 1988, ApJ, 332, 199

\bibitem[]{} Schlegel, D., Finkbeiner, D., Davis, M., ApJ, 1998, 500, 525.

\bibitem[]{} Slowikowska, A., Kuiper, L., Hermsen, W., 2005, A\&A, 434, 1097

\bibitem[]{} Standish, E.M., 1982 A\&A, 114 ,297

\bibitem[]{} Sun, X., Tr\"umper, J., Dennerl, K., Becker, W., 1993, IAU circ.~5895

\bibitem[]{} Tang, A. P. S., \& Cheng, K. S. 2001, ApJ, 549, 1039

\bibitem[]{} Tepedelenlioglu, E., \"Ogelman, H.B., 2005, ApJ, submitted (astro-ph/0505461)

\bibitem[]{} Tsuruta, S., 1998, PhR, 292, 1

\bibitem[]{}  Wang, D., Halpern, J.P., 1997, ApJ, 482, L159

  \bibitem{}{} Wang, Q. D., Li, Z. Y., \& Begelman M. C. 1993, \nat, 364, 127

\bibitem[]{}  Willingale, R., Aschenbach, B., Griffiths, R. G., Sembay, S., Warwick, R. S.,
  Becker, W., Abbey, A. F., Bonnet-Bidaud, J.-M., 2001, A\&A, 365, L212

\bibitem[]{} Yadigaroglu, I.-A., \& Romani, R. W. 1995, ApJ, 449, 211


\bibitem[]{} Yakovlev, D.G., Levenfish, K.P., Shibanov, Yu.A. 1999, Physics-Uspekhi, 169, 825

\bibitem[]{} Yancopoulos, S., Hamilton, T.T., Helfand, D.J., 1994, ApJ, 429, 832


\bibitem[]{} Zavlin, V.E., Pavlov, G.G., 2004, ApJ, 616, 452

\bibitem[]{}  Zickgraf, F.-J., Krautter, J., Reffert, S., Alcala, J.M., Mujica, R., Covino, E., Sterzik, M.F., 2005, A\&A, 433, 151

\bibitem[]{} Zhang, B., Harding, A.K., 2000, ApJ, 532, 1150

\end{thebibliography}
\end{document}